\documentclass{article}

\usepackage{amsmath}
\usepackage{amsfonts}

\numberwithin{equation}{section}

\def\doubleset#1#2{\bgroup%
\def\doit#1#2{%
\setbox\dblsetbox=\hbox{$\cstyle #1$}%
\raise#2\ht\dblsetbox\copy\dblsetbox%
\hskip-\wd\dblsetbox%
\raise-#2\ht\dblsetbox\box\dblsetbox}%
\mathchoice%
{\def\cstyle{\displaystyle}\doit#1#2}%
{\def\cstyle{\textstyle}\doit#1#2}%
{\def\cstyle{\scriptstyle}\doit#1#2}%
{\def\cstyle{\scriptscriptstyle}\doit#1#2}\egroup}
\newbox\dblsetbox

\newcommand{\plpl}{\doubleset{+}{.185}}
\newcommand{\mimi}{=}
\newcommand{\rd}{\mathrm{d}}
\newcommand{\intd}[1]{\int\mspace{-6mu}\rd#1\,}
\newcommand{\ints}[2]{\int_{#1}\mspace{-6mu}#2\,}
\newcommand{\Nabla}{\Tilde{\nabla}}
\newcommand{\bR}{\mathbb{R}}
\newcommand{\bC}{\mathbb{C}}

\begin{document}

\thispagestyle{empty}

\begin{center}
  \begin{large}
    \vspace{.7cm}

    \textbf{Supersymmetric Gauge Theories,  Vortices  and  Equivariant
    Cohomology}

    \vspace{2cm}

    W. Machin and G. Papadopoulos

    \vspace{1cm}

    \textit{Department of Mathematics \\
      King's College London \\
      Strand \\
      London WC2R 2LS }
    \vspace{2cm}
  \end{large}

\begin{abstract}
  We construct actions for $(p,0)$- and $(p,1)$- supersymmetric,
  $1\leq p\leq 4$, two-dimensional gauge theories coupled to
  non-linear sigma model matter with a Wess-Zumino term. We derive the
  scalar potential for a large class of these models. We then show
  that the Euclidean actions of the $(2,0)$ and $(4,0)$-supersymmetric
  models without Wess-Zumino terms are bounded by topological charges
  which involve the equivariant extensions of the K\"ahler forms of
  the sigma model target spaces evaluated on the two-dimensional
  spacetime. We give similar bounds for Euclidean actions of
  appropriate gauge theories coupled to non-linear sigma model matter
  in higher spacetime dimensions which now involve the equivariant
  extensions of the K\"ahler forms of the sigma model target spaces
  and the second Chern character of gauge fields. The BPS
  configurations are generalisations of abelian and non-abelian
  vortices.
\end{abstract}

\end{center}

\newpage

\section{Introduction}

Two-dimensional field theories have found many applications in string
theory.  For example, two-dimensional sigma models and some
two-dimensional gauge theories have been used to model the dynamics of
fundamental and D-strings, respectively.  The small fluctuations of
strings which arise as intersections in various brane configurations
are described by two-dimensional gauge theories coupled to scalars.
Because of this, many of the properties and the various objects that
arise in gauge theories coupled to scalars have a brane interpretation
\cite{strom, gppkt, hanwit, townsend}.  Supersymmetric gauge theories
coupled to linear sigma models have been constructed in \cite{witten}
and they have been used to illuminate the relation between
Landau-Ginzburg models and Calabi-Yau spaces.  Recently a
two-dimensional gauged theory coupled to a linear sigma model was used
to investigate aspects of the dynamics of vortices using branes
\cite{tong}.

In two dimensions, the Wess-Zumino term has the same mass dimension as
the kinetic term of sigma model scalars. Therefore two-dimensional
supersymmetric gauged theories can couple to non-linear sigma model
matter which also has a non-vanishing Wess-Zumino coupling. Such a
theory is renormalizable.  The gauging of supersymmetric
two-dimensional non-linear sigma models with a Wess-Zumino term has
been considered in \cite{hs, hgps}. However in these papers the part
of the action which involves the gauge field kinetic terms has not
been given.  It has been found in \cite{hs} that the Wess-Zumino term
of a non-linear sigma model cannot always be gauged. The conditions
for gauging a Wess-Zumino term have been identified as the
obstructions to the extension of the closed form associated with the
Wess-Zumino term to an element of the equivariant cohomology \cite{ab}
of the sigma model target space \cite{jffss}.  Scalar potentials for
supersymmetric two-dimensional sigma models with Wess-Zumino term have
been investigated in \cite{hgpt, gppt,gppt2,gppt3}.

In this paper we shall construct the actions of (p,0)- and
(p,1)-supersymmetric, $1\leq p\leq 4$, two-dimensional gauge theories
coupled to non-linear sigma model matter and with non-vanishing
Wess-Zumino term. In addition we shall also consider the scalar
potentials that arise in these theories.  This will generalise
various partial results that  have already appeared in the literature.
To simplify the description of the results from here on, we shall use
the term \lq sigma models' instead of the term \lq non-linear sigma
models' unless otherwise explicitly stated.  The method we shall use
to construct the various actions of supersymmetric two-dimensional
gauged theories coupled to sigma models is based on the superfields
found in the context of supersymmetric sigma models \cite{phgp, phgpb}
and later used in the context supersymmetric gauged sigma models
\cite{hgps}. One advantage of this method is that it keeps manifest
the various geometric properties of the couplings that appear in these
theories. This will be used in the second part of the paper to
construct of various bounds for vortices.  Since the parts of the
actions that we shall describe involving the kinetic term of the sigma
model scalars, the Wess-Zumino term and their couplings to gauge
fields are known, we shall focus on the kinetic term of the gauge
fields and the scalar potentials of these theories. We shall allow the
gauge couplings to depend on the sigma model scalars and we shall
derive the various conditions on these couplings required by gauge
invariance and supersymmetry.  We shall find that the scalar potential
of gauge theories coupled to sigma models in two dimensions, even in
the presence of Wess-Zumino term, is the sum of a `F' term or a `D'
term or both.  The presence of a D-term, or Fayet-Iliopoulos term, may
come as a surprise. This is because in the presence of a Wess-Zumino
term the geometry of the target space, say, of (2,0)-supersymmetric
models where such a term is expected, is {\it not} K\"ahler and but
K\"ahler with torsion (KT). Thus the K\"ahler form is not closed and
there are no obvious moment maps. However, it has been shown in
\cite{ggpp} that KT geometries under certain conditions admit moment
maps and are those that appear in the Fayet-Iliopoulos term of these
models. Similar results hold for the (4,0)-supersymmetric models.  We
shall observe that the gauged (p,1), $p=1,2,4$, multiplets are
associated with scalar superfields. For the gauge theories with (2,1)
and (4,1) supersymmetry, these scalar multiplets satisfy the same
supersymmetry constraints as the associated sigma model multiplets.
Therefore these gauge theories can be thought of as sigma models with
target spaces ${\cal L}G\otimes \bR^p$, $p=1,2,4$.  This will allow us
to combine the (p,1) gauge multiplet and the standard sigma model
$(p,1)$ multiplet to a new sigma model multiplet. As a result, sigma
model type of actions can be written for these gauge theories coupled
to matter for which the associated couplings depend on the scalar
fields of both the sigma model and gauge multiplets. This generalizes
the results of \cite{hgps}.

In the second part of this paper, we shall show that the Euclidean
actions of (2,0)- and (4,0)-supersymmetric two-dimensional gauge
theories coupled to sigma models with a Fayet-Iliopoulos term but with
{\it vanishing} Wess-Zumino term admit bounds. In particular we shall
find that the Euclidean action $S_E$ of the (2,0)-supersymmetric
theory is bounded by the absolute value of a topological charge ${\cal
  Q}$ which is the integral over the two-dimensional spacetime of the
{\it equivariant extension of the K\"ahler form} of its sigma model
target space, $S_E\geq |{\cal Q}|$.  The sigma model manifold in the
(4,0)-supersymmetric theory is hyper-K\"ahler and so there are three
K\"ahler forms each having an equivariant extension. The Euclidean
action $S_E$ of the (4,0)-supersymmetric theory is bounded by the
length of the three topological charges ${\cal Q}_1, {\cal Q}_2, {\cal
  Q}_3$ each associated with the integral over the two-dimensional
spacetime of the equivariant extensions of the three K\"ahler forms,
$S_E\geq \sqrt {{\cal Q}^2_1+ {\cal Q}^2_2+ {\cal Q}^2_3}$.  This is
another application of the equivariant cohomology in the context of
two-dimensional gauged sigma models which is distinct from that found
in \cite{jffss} and we have mentioned above.  (For many other
applications see for example \cite{moore}.)  The configurations that
saturate these bounds are vortices and include the Nielsen-Olesen type
of vortices \cite{hnpo} associated with gauge theories coupled to
linear sigma models.  In particular the bounds above generalise that
found by Bogomol'nyi in \cite{bog} for the abelian vortices of a gauge
theory coupled to a single linear complex scalar field.

We also find that similar bounds exist in higher dimensions for action
type of functionals that involve maps between  K\"ahler manifolds of
any dimension coupled to gauge fields or maps from a K\"ahler manifold
into a hyper-K\"ahler manifold again coupled to gauge fields. The
structure of these functionals is such that it  includes the Euclidean
actions of some supersymmetric gauge theories in higher dimensions
coupled to sigma models with Fayet-Iliopoulos terms.  In particular
the first case, which involves maps between two-K\"ahler spaces,
includes the Euclidean action of a four-dimensional $N=1$
supersymmetric gauge theory coupled to a sigma model. The latter case,
which involved maps from a K\"ahler manifold into a hyper-K\"ahler one,
can be associated with the Euclidean action of a four-dimensional
$N=2$ supersymmetric gauge theory coupled to a sigma model. Note that
in $N=1$ theories in four dimensions the sigma model target space is
K\"ahler while in the $N=2$ theories in four dimensions the sigma
model target space is hyper-K\"ahler.  In all these cases the action
functionals are bounded by {\it topological charges} which involve the
{\it equivariant extensions} of the K\"ahler forms of the sigma model
target space as well as the second Chern character of the gauge
fields.  Our results are different from those of \cite{matha, mathb}
for non-abelian vortices which involve non-abelian gauge theories
coupled to linear sigma model matter. Note that in the bound
constructed in \cite{matha}, the topological term involves the first
class and second Chern character of the gauge fields instead of the
equivariant extension of the K\"ahler form and the second Chern
character of the gauge field that we find. It turns out that in the
case of gauge theories coupled to {\it linear} sigma models of
\cite{witten} the two different topological charges can be related,
see also \cite{taubes}. However this involves a partial integration
procedure in which various surface terms are taken to vanish. We
remark that in the construction of the bounds that involve the
equivariant extensions of the K\"ahler forms, and also in \cite{bog},
the topological terms are identified with what remains after writing
the Euclidean actions of the theories as a sum of squares without the
use of partial integrations.  It is clear that our results can also be
used to construct bounds for solitons in appropriate gauge theories
coupled to sigma model matter in odd-spacetime dimensions. This is the
usual situation where instantons in a n-dimensional theory can be
thought of as static solitons of a (n+1)-dimensional theory. In
particular, there is a bound for the energy of static configurations
of a three-dimensional $N=2$ supersymmetric gauge theory coupled to
sigma model matter. This new bound is an extension to gauged theories
of the of bounds found in \cite{apt} and generalizes that of
\cite{rebbi}.

This paper is organised as follows: In section two, we shall describe
non-supersymmetric gauge theories coupled to sigma model matter to set
up our notation and give some universal conditions which appear in all
such models. In section three, we shall describe the
(1,0)-supersymmetric gauge theories coupled to sigma model matter with
a Wess-Zumino term. We shall also give the scalar potential of the
system.  In sections four and five, we shall describe the (2,0)- and
(4,0)-supersymmetric gauge theories coupled to sigma model matter and
give the scalar potentials of the systems. We shall find that these
systems have Fayet-Iliopoulos terms constructed from the moment maps
of KT and HKT geometries, respectively. In section six, we shall
describe the (1,1)-supersymmetric gauge theory coupled to sigma model
matter. We shall find that both the gauge multiplet and the sigma
model multiplet are scalar superfields allowing for non-polynomial
interactions between them.  In sections seven and eight, we shall
describe the (2,1)- and (4,1)-supersymmetric gauge theories coupled to
sigma model matter and give the scalar potentials of the systems. In
both these theories the gauge multiplets and the sigma model
multiplets are scalar superfields and therefore admit non-polynomial
interactions.  In section nine, we describe a bound for the Euclidean
action of a (2,0)-supersymmetric gauge theory without Wess-Zumino
term.  We show that the equivariant extension of the K\"ahler class
enters in the bound. The BPS configurations are vortices and the
scalar fields take values in a curved manifold $M$.  In section ten,
we describe a bound for the Euclidean action of a (4,0)-supersymmetric
gauge theory without Wess-Zumino term.  We show that the topological
term in the bound is a linear combination equivariant extensions of
the K\"ahler classes of the hyper-K\"ahler manifold which is the sigma
model target space.  In sections eleven and twelve, we present a
generalisation of the bounds we have described for (2,0)- and
(4,0)-supersymmetric models to a class of gauge theories coupled
to scalars which  are  maps
between K\"ahler manifolds or from a K\"ahler manifold to a
hyper-K\"ahler, respectively.  The topological
term this time involves, apart from the equivariant extensions of
K\"ahler forms, the second Chern character of the gauge connection.
Finally in section thirteen, we give our conclusions.

\section{Two-dimensional gauged sigma models with Wess-Zumino term}

\subsection{Geometric Data and Action}
\label{sub:data}

To describe two-dimensional supersymmetric gauge theories coupled to sigma model
matter with a Wess-Zumino term,  it is instructive to begin with
a non-supersymmetric system as a toy example.  Let $\Xi$ be the
two-dimensional Minkowski spacetime with light-cone coordinates
$(x^\plpl,x^\mimi)$.  The fields of the model that we shall consider
here are the following: a gauge potential $A$ with gauge group $G$,
sigma model matter fields $\phi$ which are locally maps from $\Xi$ into a
sigma model manifold or target space $M$ and real fermions $\psi_-$
and $\lambda_+$ on $\Xi$ of opposite chirality.

The couplings of two-dimensional gauged sigma model are
described by a Riemannian metric $g$ on $M$ and the
Wess-Zumino term which is a locally-defined two-form
$b$ on $M$; $H=\rd b$ is a globally defined closed three-form on
$M$. In addition, the gauge group $G$ acts on $M$
 leaving invariant both the metric and the Wess-Zumino term, ie
\begin{align}
\label{eq:ghinvariant}
  \mathcal{L}_a g &=0
  & \mathcal{L}_a H&=0
\end{align}
where $\mathcal{L}_a$ is the Lie-derivative with respect to the vector
fields $\{\xi_a: a=1,\dots, \dim {\cal L}G\}$ generated by the action
of the gauge group $G$ on $M$; ${\cal L}G$ is the Lie algebra of $G$. Therefore,
 we have
\begin{align}
[ \xi_a, \xi_b ]^i =- f_{ab}^{\phantom{ab}c}\xi_c^{\phantom{c}i}\ ,
\end{align}
where $f$ are the structure constants of $G$; $a,b,c=1,\dots, \dim
{\cal L}G$ are gauge indices.  The first condition in
\eqref{eq:ghinvariant} implies that $\xi_a$ are Killing vectors, ie
\begin{align}
\label{eq:killingcnd}
  \nabla_i \xi_{aj}+\nabla_j\xi_{ai}=0
\end{align}
where $\nabla$ is the Levi-Civita connection of $g$ and
$i,j=1,\dots, \dim M$.
 The second condition in \eqref{eq:ghinvariant} together with $\rd H=0$ imply
 that $i_a H$ is closed and so
\begin{align}
\label{eq:u1form}
  \xi_a^{\phantom{a}i}H_{ijk} &= 2\partial_{\smash{[j}}w_{k]a}
\end{align}
for some locally defined one-form $w_a$.

It is useful to give a geometric interpretation for the gauge
potential $A$ of the sigma model field $\phi$.  Let $P$ be a principal
bundle over the spacetime $\Xi$ with fibre the gauge group $G$.  The
gauge potential $A$ is locally the pull-back of a connection one-form
$A$ of $P$ onto a open set of the spacetime $\Xi$. The sigma model
maps $\phi$ are sections of the bundle $P\times_G M$. Locally they can
be represented as maps from the spacetime $\Xi$ into $M$.

To describe the couplings of the fermions, we consider two vector
bundles $E$ and $F$ over $M$ equipped with connections $B$ and $C$ and
with fibre metrics $h$ and $k$, respectively. The fermions $\psi_-$
and $\lambda_+$ can be thought of as sections of $S_-\otimes E$ and
$S_+\otimes F$, respectively, where $S_-$ and $S_+$ are spin bundles
over $\Xi$ associated to the two inequivalent real representation of
$Spin(1,1)$. Note that in two dimensions there are Majorana-Weyl
fermions and so two inequivalent one-dimensional real spinor
representations of $Spin(1,1)$.  In addition we shall assume that the
connections $B$ and $C$ as well as the fibre metrics $h$ and $k$ are
invariant under the action of the gauge group on $M$. These conditions
imply that
\begin{align}
\label{eq:bgaugecnd}
 \mathcal{L}_a B_{i\phantom{A}B}^{\phantom{i}A}&=
 -\nabla_i U_{a\phantom{A}B}^{\phantom{a}A}\\
\label{eq:hgaugecnd}
  \mathcal{L}_a h_{AB}&=-U_a{}^C{}_A h_{CB}-U_a{}^C{}_B h_{AC}
  \ ,
\end{align}
where
\begin{align}
  \nabla_i U_{a\phantom{A}B}^{\phantom{a}A} &
  =\partial_i U_{a\phantom{A}B}^{\phantom{a}A}
   +  B_i{}^A{}_C U_a{}^C{}_B
- U_a{}^A{}_C B_i{}^C{}_B \ ,
\end{align}
and $U_a$ are infinitesimal gauge transformations, $A,B,C=1,\dots, {\rm rank}\,E$,
and similarly for the connection $C$ and fibre metric $k$. The
above conditions on the connection $B$ have appeared in \cite{hull}.
An action for the  fields $A$, $\phi$, $\psi_-$ and $\lambda_+$ is
\begin{align}
\label{eq:gauge00action}
\begin{split}
  S ={}& \intd{^2x} \bigl(u_{ab} F^a_{\plpl\mimi} F^b_{\plpl\mimi}+
       g_{ij} \nabla_\plpl \phi^i \nabla_\mimi \phi^j
       - V(\phi)
       \bigr)\\
    & + \intd{^2x\rd t}\bigl(
        H_{ijk}\partial_t \phi^i \nabla_\plpl \phi^j
    \nabla_\mimi\phi^k - w_{ia}\partial_t\phi^i F_{\plpl\mimi}^a
    \bigr)\\
    & +\intd{^2x}(ih_{AB}\psi_-^A\Nabla_\plpl\psi_-^B-ik_{A'B'}
     \lambda_+^{A'}\Nabla_\mimi \lambda_+^{B'})
\end{split}
\end{align}
where $u_{ab}=u_{ab}(\phi)$ are the gauge couplings which in general
depend on $\phi$, $V$ is a scalar potential
and
\begin{align}
  F_{\plpl\mimi}=[\nabla_\plpl,\nabla_\mimi] ~,
\end{align}
where we have suppressed gauged indices.
The covariant derivatives in the action above are defined as follows,
 \begin{align}
  \nabla_\mu \phi^i=\partial_\mu \phi^i+ A^a_\mu \xi_a^i
\end{align}
and
\begin{align}
 \Nabla_\mu\psi_-^A &= \partial_\mu\psi_-^A
  + \nabla_\mu\phi^iB_{i\phantom{A}B}^{\phantom{i}A}\psi_-^B
  + A_\mu^a U_{a\phantom{A}B}^{\phantom{a}A}\psi^B  \ ,
\end{align}
$\mu=\plpl, \mimi$, and similarly for $\Nabla \lambda_+^{A'}$.
The latter can be rewritten as
\begin{align}\label{eq:covdev}
 \Nabla_\mu\psi_-^A &= \partial_\mu\psi_-^A
  + \partial_\mu\phi^iB_{i\phantom{A}B}^{\phantom{i}A}\psi_-^B
  + A_\mu^a \mu_{a\phantom{A}B}^{\phantom{a}A}\psi^B  ~.
\end{align}
where
\begin{align}
\mu_a{}^A{}_B=U_a{}^A{}_B+\xi_a{}^i B_i{}^A{}_B ~.
\end{align}

Observe that the part of the action involving the Wess-Zumino term
has been written as an integral over a three-dimensional space.
The conditions for this term to be written in a two-dimensional form
as well as the conditions for the gauge invariance of the action
will be investigated in the next section, see also \cite{hs}.

\subsection{Conditions for gauge invariance}
\label{sub:gaugecon}

The gauge transformations of the fields are
\begin{align}\begin{split}
  \delta_\epsilon A_\mu^a &=- \nabla_\mu \epsilon^a \\
  \delta_\epsilon \phi^i &= \epsilon^a \xi_a^i \\
  \delta_\epsilon \psi_-^A &= \epsilon^a U_{a\phantom{A}B}^{\phantom{a}A} \psi^B_- \\
  \delta_\epsilon \lambda_+^{A'}&=\epsilon^a V_{a\phantom{A'}B'}^{\phantom{a}A'}
   \lambda_+^{B'}
\end{split}\end{align}
where $\epsilon$ is the parameter of infinitesimal gauge
transformations.  Some of the conditions required for the
invariance of the action~\eqref{eq:gauge00action} have been
incorporated as part of the geometric data of the sigma model
in the previous section. In particular, in addition to the conditions
\eqref{eq:ghinvariant} and \eqref{eq:bgaugecnd}, we require that (i) $w_a$
is a globally defined one-form on $M$ which (ii) satisfies
\begin{align}
{\cal L}_a w_b=-f_{ab}{}^c w_c\ .
\label{giw}
\end{align}

To write the Wess-Zumino part of the action in a two-dimensional form, it
is necessary for the relevant three form to be closed. This in addition
requires that
\begin{align}
\xi^i_a w_{ib}+\xi^i_b w_{ia}=0\ .
\label{weq}
\end{align}
Then  the three-dimensional part of the action \eqref{eq:gauge00action}
can be written locally \cite{hs} as
\begin{align}
  \begin{split}
    S' = \intd{^2x} \bigl(&
     b_{ij} \partial_\plpl\phi^i \partial_\mimi\phi^j
   - A_\plpl^a w_{ia} \nabla_\mimi\phi^i\\
&    + A_\mimi^a w_{ia} \nabla_\plpl\phi^i
    + A_\plpl^a A_\mimi^b \xi_b^{\phantom{b}i}w_{ai}
    \bigr)\ .
  \end{split}
\end{align}
The conditions above that require (i) $H$ to be invariant under the
group action, (ii) $w_a$ to be globally defined on $M$, (iii)
\eqref{giw} and (iv) \eqref{weq} are those for the closed form $H$ to
have an extension, i.e.\  equivariant extension \cite{ab}, as a closed
form in $M\times_G EG$ \cite{jffss}.

Next let us consider the conditions for gauge invariance
of part of the action~\eqref{eq:gauge00action} involving the fermions.
We find that this requires that
 \begin{align}
  \mathcal{L}_c u_{ab} + u_{db}f^d_{\phantom{d}ca}+u_{ad}f^d_{\phantom{d}cb}=0
\end{align}
for the gauge couplings $u_{ab}$,
\begin{align}
\label{eq:psigaugecnd}
 \mathcal{L}_aU_b-\mathcal{L}_b U_a = [U_a,U_b] 
  -f_{ab}^{\phantom{ab}c}U_c
\end{align}
 and
\begin{align}
  \mathcal{L}_a V&=0 ~.
\end{align}
We have not assumed that $\nabla_i h = 0$. However given a connection
on a vector bundle with a fibre metric $h$, there always exist another
connection $\nabla'$ such that $\nabla_i'h=0$. Suppose that the
$\nabla'$ is used for the fermionic couplings. If this is the case,
the gauge group is $O(N)$ and therefore the right-hand-side
of~\eqref{eq:hgaugecnd} vanishes. 

Observe that the equation~\eqref{eq:bgaugecnd} can be written in a
more covariant form as
\begin{align}
\xi_a^j G_{ij\phantom{A}B}^{\phantom{ij}A}
= \nabla_i \mu_{a\phantom{A}B}^{\phantom{a}A} ~.
\end{align}

In what follows we shall assume that the conditions stated in this
section by requiring gauge invariance of the non-supersymmetric model
described by the action \eqref{eq:gauge00action} hold. We shall see
that for supersymmetric sigma models more conditions are necessary.

\section{$(1,0)$ supersymmetric gauged models}

The (1,0)-supersymmetric gauged sigma model involves the coupling of
three different (1,0)-multiplets. To simplify the construction of this
model we shall describe each multiplet and the conditions for
supersymmetry and gauge invariance separately. There are different
ways of approaching this problem. Here we shall use `standard'
(1,0)-superfields. The action will be constructed using
(1,0)-superspace methods.

\subsection{The gauge multiplet}

The (1,0)-superspace $\Xi^{1,0}$ has coordinates
$(x^\plpl,x^\mimi,\theta^+)$, where $(x^\plpl,x^\mimi)$ are bosonic
light-cone coordinates and $\theta^+$ is a Grassmann odd-coordinate.
The (1,0)-supersymmetric Yang-Mills multiplet with gauge group $G$ is
described by a connection $A$ in superspace which has components
$(A_\plpl, A_\mimi, A_+)$.  In addition, it is required that these
satisfy the supersymmetry constraints \cite{hgps}
\begin{gather}
  [ \nabla_+, \nabla_+ ] = 2i\nabla_\plpl
\mspace{100mu} [ \nabla_\plpl, \nabla_\mimi ] = F_{\plpl\mimi}     
\nonumber\\
[ \nabla_+, \nabla_\mimi ] = W_- \ ,
\end{gather}
where $F_{\plpl\mimi}$, $W_-$ are the components of the curvature of the
superspace connection $A$. (The gauge indices have been suppressed.)
Jacobi identities imply that
\begin{align}
  \nabla_+ W_- &= iF_{\plpl\mimi}
\end{align}
Therefore the independent components of the gauge multiplet are
\begin{align}
  \chi^a &= W_-| & F^a_{\plpl\mimi} &=
  -i\nabla_+W^a_-|
\end{align}
where $\chi^a$ is the gaugino and $F^a$ is the two-form gauge field
strength and the vertical line denotes evaluation of the associated
superfield at $\theta^+=0$. This notation for identifying
the components of a superfield will also be used
later for other theories.

\subsection{Sigma model multiplets}

To described the sigma model  multiplet that couples to the
above gauge field, we introduce a Riemannian manifold $M$ with metric
$g$ and a locally defined two form $b$. In addition we assume that $M$
admits a vector bundle $E$ with fibre metric $h$, connection $B$ and a section $s$.
The data required for the description of the sigma model multiplet are the same
as those given in section \ref{sub:data}. In addition
we take the section $s$ to satisfy
 \begin{align}
 \label{eq:sgaugecnd}
  \mathcal{L}_a s_A = -U^{\phantom{a}B}_{a\phantom{B}A} s_B\ .
  \end{align}

The $(1,0)$-supersymmetric sigma model multiplet is described by a
real scalar superfield $\phi$ and a fermionic {\sl superfield} $\psi_-$. The
superfield $\phi$ is a map from the superspace $\Xi^{1,0}$ into a
sigma model manifold $M$ and the fermionic superfield $\psi_-$ is a
section of the bundle $\phi^*E\otimes S_-$; $S_-$ is a spin bundle
over $\Xi^{1,0}$.

The components of the superfields $\phi$, $\psi_-$ are
\begin{align}
  \phi^i &= \phi^i|
  & \lambda^i_+ &= \nabla_+\phi^i|  \nonumber\\
  \psi_-^A &= \psi_-^A| & \ell^A &=
  \Nabla_+\psi_-^A| \ ,
\end{align}
where the covariant derivatives are defined as
\begin{align}
  \nabla_+\phi^i &= D_+\phi^i+A_+^a \xi_a^i \nonumber\\
  \Nabla_+\psi_-^A &= D_+\psi_-^A + \nabla_+\phi^i
  B_{i\phantom{A}B}^{\phantom{i}A} \psi_-^B + A_+^a
  U_{a\phantom{A}B}^{\phantom{a}A} \psi_-^B ~,
\end{align}
where $D_+$ is the usual flat superspace derivative, $D_+^2=i\partial_\plpl$.

\subsection{Supersymmetric action}

It is straightforward to couple the gauge multiplet to
$(1,0)$-supersymmetric sigma model matter. The full action is
\begin{align}
\label{eq:01ta}
  S &= S_g +S_\sigma+ S_f + S_p
\end{align}
where
\begin{align}
  S_g &=- \intd{^2x\rd\theta^+}\bigl( u_{ab}W_-^a\nabla_+W_-^b -i z_a W^a_-\bigr) \ ,
\end{align}
where $u_{ab}=u_{ab}(\phi)$, $u_{ab}$ is not necessarily symmetric in
the gauge indices and $z_a$ is a theta type of term which may depend
on the scalar field $\phi$, $z_a=z_a(\phi)$.

The gauge covariant supersymmetric action for the fields $\phi$
is~\cite{hgps}
\begin{align}
\begin{split}
  S_\sigma ={}&-i \intd{^2x\rd\theta^+}
  g_{ij}\nabla_+\phi^i\nabla_\mimi\phi^j \\
  &-i \intd{^2x\rd t \rd\theta^+}\bigl(
  H_{ijk}\partial_t\phi^i\nabla_+\phi^j\nabla_{\mimi}\phi^k -
  w_{ia}\partial_t\phi^i W_-^a \bigr)
\end{split}
\end{align}
which as in section \ref{sub:data} can also be written as an integral
over $\Xi^{1,0}$ superspace provided
that $H$ admits an equivariant extension. In particular we have
\begin{align}
\label{eq:gauge10action}
\begin{split}
  S_\sigma = -i\intd{^2x\rd\theta^+}\bigl(&
    g_{ij}\nabla_+\phi^i\nabla_\mimi\phi^j
    + b_{ij} D_+\phi^i \partial_{\mimi}\phi^j \\
&   - A_+^a w_{ia} \partial_\mimi\phi^i
    +A_\mimi^a w_{ia} D_+\phi^i
    + A_+^a A_\mimi^b \xi^i_{[b} w_{a]i}
    \bigr)~.
\end{split}
\end{align}

The action of the gauged fermionic multiplet  is~\cite{hull}
\begin{align}
  S_f &= \intd{^2x\rd\theta^+}h_{AB}\psi_-^A\Nabla_+\psi_-^B~.
\end{align}
The definition of the covariant derivative $\Nabla_+$ is similar to the
one given in section \eqref{sub:data} for the covariant derivative
$\Nabla_\plpl$.

The action for the potential term is
\begin{align}
  S_p &= \intd{^2x\rd\theta^+} m s_A \psi_-^A
\end{align}
which is similar to that of the ungauged model in~\cite{hgpt}.

The superfields transform under the gauge group $G$ as
\begin{align}
\delta A_\mu&=-\nabla_\mu\epsilon^a
  & \delta\phi^i &= \epsilon^a\xi_a^{\phantom{a}i}(\phi)
& \delta\psi_-^A &= \epsilon^a U_{a\phantom{A}B}^{\phantom{a}A} \psi_-^B ~.
\end{align}
where $\mu=\plpl, \mimi, +$ is a $\Xi^{1,0}$ superspace index and $\epsilon^a$ is an infinitesimal
gauge transformation parameter.
 Gauge invariance of the action \eqref{eq:01ta} requires, in addition
to the conditions given in section \eqref{sub:gaugecon}, the condition
\eqref{eq:sgaugecnd} and
\begin{align}
{\cal L}_a z_b=-f_{ab}{}^c z_c~.
\end{align}

\subsection{The action of (1,0)-model in components and scalar potential}

The action of (1,0)-supersymmetric two-dimensional gauged sigma model
described by the action \eqref{eq:01ta} can be easily written
in components by performing the $\theta^+$ integration and using the
definition of the various component fields of the (1,0)-multiplets which
we have described in the previous sections.
In particular we find for the part of the action involving the
kinetic term of the gauge multiplet that
\begin{align}
\begin{split}
  S_{g}=\intd{^2x}\bigl(&
  u_{ab}F_{\plpl\mimi}^aF_{\plpl\mimi}^b+z_a F_{\plpl\mimi}^a \\
  &-i
  u_{ab}\chi_-^a\nabla_{\plpl}\chi_-^b
  +i\partial_iu_{ab} \lambda^i_+
 \chi_-^a F_{\plpl\mimi}^b +i\partial_i z_a \lambda_+^i \chi_-^a
   \bigr)\ .   
\end{split}
\end{align}
Next we find that
\begin{align}
  \begin{split}
    S_\sigma = \intd{^2x} \bigl(&
    g_{ij} \nabla_{\plpl}\phi^i \nabla_\mimi\phi^j
    + b_{ij} \partial_\plpl\phi^i \partial_\mimi\phi^j
      +i \lambda_+^i \Nabla^{(+)}_\mimi \lambda_+^j
    \\
&   +i w_{ia} \lambda_+^i \chi_-^a
    -  A_\plpl^a w_{ia} \partial_\mimi\phi^i
    +  A_\mimi^a w_{ia} \partial_\plpl\phi^i \\
&   -  A_\plpl^a A_\mimi^b \xi_{[b}^{\phantom{[b}i} w_{a]i}
    \bigr)~,
  \end{split}
\end{align}
where $\nabla^{(\pm)}$ are the usual metric connections with torsion
$\pm H$ and $\Nabla^{(\pm)}$ are the associated connections
involving also the gauge connection $A$.  For the fermionic multiplet
we have
\begin{align}
  S_f = \intd{^2x} \bigl(
      -i h_{AB} \psi_-^A \Nabla_\plpl \psi_-^B
      + h_{AB} \ell^A \ell^B
      - \frac{1}{2} G_{ijab} \psi_-^A \psi_-^B \lambda_+^i \lambda_+^j
      \bigr)
\end{align}
and
\begin{align}
S_p = \intd{^2x} \bigl(
      \nabla_i s_A \lambda_+^i
      + s_A \ell^A
      \bigr)\ .
\end{align}

The scalar potential in these models is precisely that of  the ungauged
(1,0) sigma models in \cite{hgpt}, ie.
\begin{align}
  V=\frac{1}{4}m^2 h^{AB} s_A s_B~ .
\end{align}
So we find that only `$F$-terms' contribute to the potential. This is
because the gauge multiplet does not have an auxiliary field.
Therefore the classical vacua of the theory are the points of the
sigma model manifold $M$ for which the section $s$ vanishes modulo gauge
transformations.  Therefore the vacua of the theory are those
orbits of the gauge group $G$ in $M$ for which the
section $s$ vanishes.

\section{$(2,0)$ supersymmetry}

\subsection{The gauge multiplet}

The $(2,0)$ superspace $\Xi^{2,0}$ has coordinates
$(x^\plpl,x^\mimi,\theta_{0}^+,\theta_{1}^+)$ where $(x^\plpl,x^\mimi)$
are the usual light-cone coordinates and $\{\theta_{p}^+: p=0,1\}$ are
anticommuting  coordinates. The (2,0)-supersymmetric Yang-Mills
multiplet is described by a connection $A$ in $\Xi^{2,0}$ superspace with
components $(A_\plpl,A_\mimi,A_{p+})$, $p=0,1$. In addition it is
required that these satisfy the supersymmetry constraints \cite{hgps}
\begin{align}
  [ \nabla_{p+},  \nabla_{q+} ] &= 2i\delta_{pq}\nabla_\plpl
& [ \nabla_\plpl, \nabla_\mimi ] &= F_{\plpl\mimi}
& [ \nabla_{p+},  \nabla_\mimi ] &= W_{p-} \ ,
\end{align}
where $p,q=0,1$.
Jacobi identities imply that
\begin{align}
  \nabla_{p+} W_{q-} + \nabla_{q+} W_{p-}
&=  2i\delta_{pq}F_{\plpl\mimi}
\end{align}
The components of the gauge multiplet are
\begin{align}
  \chi_{0-} &= W_{0-}|
&\chi_{1-} &= W_{1-}|  \nonumber\\
i F_{\plpl\mimi} &= \nabla_{0+}W_{0-}|
& f &= \nabla_{0+}W_{1-}| ~.
\end{align}
The components, $(\chi_{0-}, \chi_{1-})$, are the gaugini which are real
chiral fermions in two dimensions, $F_{\plpl\mimi}$ is the field
strength and $f$ is an auxiliary field. (We have suppressed the gauge indices.)

\subsection{The sigma model multiplet}

It is well known the target manifold $M$ of (2,0)-supersymmetric
  ungauged sigma model
  is K\"ahler  manifold with torsion (KT).
Therefore $M$ is a hermitian manifold with metric $g$ and equipped
with a complex structure $J$ which is parallel 
with respect to the $\nabla^{(+)}$
connection.  This connection is a metric
connection with torsion $H$, ie $\nabla^{(+)}=\nabla+{1\over2}H$ where
$\nabla$ is the Levi-Civita connection.
 (For the definition of these geometries see \cite{phgpt, ggks}).
 To gauge the model,  we  assume as in section \ref{sub:data}
 that the gauge group $G$ acts on $M$
 and leaving invariant the metric $g$ and the Wess-Zumino term $H$. In addition
 we require that the action of the group $G$ is holomorphic. This means
 that
\begin{align}
  \mathcal{L}_a J=0 \ ,
\end{align}
where the Lie derivative is along vector fields $\xi_a$ generated by
the group action of $G$.  The sigma model fields are
maps $\phi:\Xi^{2,0}\rightarrow M$ into a complex manifold $M$ which
in addition satisfy
\begin{align}
\nabla_{1+}\phi^i= J^i{}_j \nabla_{0+}\phi^j \ ,
\end{align}
where $\nabla_{p+}\phi=D_{p+}\phi^i+ A_{p+}^a \xi_a^i$, $p=0,1$.  Note that the
requirement for $M$ to be a complex manifold can be derived from the
above condition.

The components of the sigma model multiplet $\phi$ are as follows:
\begin{align}
\phi^i&=\phi^i| & \lambda^i_{+}=&\nabla_{0+}\phi^i|~.
\end{align}

\subsection{The fermionic multiplet}

Let $E$ be a vector bundle over $M$ equipped with a connection $B$
 and a fibre (almost)
 complex structure $I$.
The fermionic multiplet $\psi_-$ is a section of $\phi^*E\otimes S_-$
over the $\Xi^{2,0}$, where $S_-$ is a spin bundle over $\Xi^{2,0}$.
In addition we require that the fermionic multiplet $\psi_-$ satisfies
\begin{align}
  \Nabla_{1+}\psi_-^A= I^A{}_B \Nabla_{0+}\psi_-^B+ \frac{1}{2}m L^A\ ,
  \label{con20con}
\end{align}
where $L$ is a section of $E$
and
\begin{align}
  \Nabla_{p+}\psi_-^A=
  D_{p+}\psi^A+\nabla_{p+}\phi^i B_i{}^A{}_B \psi_-^B+ A^a_{p+}
  U_a{}^A{}_B  ~,
\end{align}
for $p=0,1$.

Compatibility of the  condition \eqref{con20con} with gauge transformations
requires that
\begin{align}\begin{split}
{\cal L}_a I^A{}_B&= U_a{}^A{}_C I^C{}_B-I^A{}_C U_a{}^C{}_B
\\
{\cal L}_a L^A&=  U_a{}^A{}_B L^B\ .
\end{split}\end{align}
These are the conditions for the gauge transformations and the (2,0)-supersymmetry
transformations to commute.

The compatibility of the condition \eqref{con20con}
 with the algebra of covariant
derivatives $\nabla$ implies the following conditions:
\begin{align}\begin{split}
G_{kl\phantom{A}B}^{\phantom{kl}A} J^i{}_k J^l{}_j&=G_{ij\phantom{A}B}^{\phantom{ij}A}
\\
J^k{}_i\nabla_k L^A- I^A{}_B \nabla_iL^B&=0
\\
J^k{}_i\nabla_k I^A{}_B- I^A{}_C\nabla_iI^C{}_B&=0~.
\label{susyalcon}
\end{split}\end{align}
These are precisely the conditions required for the off-shell closure of
(2,0) supersymmetry algebra.

It is always possible to find a connection $B$ on the bundle $E$ such
that $\nabla I=0$. In such case the last condition in
\eqref{susyalcon} is satisfied.  Decomposing $E\otimes \bC$ as
$E\otimes \bC={\cal E}\oplus \bar {\cal E}$ using $I$, the first
condition implies that ${\cal E}$ is a holomorphic vector bundle.
Then the second condition in \eqref{susyalcon} implies that the
section $L$ is the real part of a holomorphic section of ${\cal E}$.

The components of the fermionic multiplet are as follows:
\begin{align}
\psi_-^A&=\psi_-^A| & \ell^A=&\Nabla_{0+}\psi_-^A| \ ,
\end{align}
where $\psi_-$ is a two-dimensional real chiral fermion and $\ell$ is
an auxiliary field.

\subsection{Action}

The action of the (2,0)-supersymmetric gauged sigma model can be written
as
\begin{align}
  S=S_g+S_\sigma+S_f  \ ,
\end{align}
where $S_g$ is the action of the gauge multiplet, $S_\sigma$ is the
action of the sigma-model multiplet and $S_f$ is the action of the
fermionic multiplet. We shall describe each term separately.

\subsection{The gauge multiplet action}

The most general action for the (2,0)-supersymmetric gauge multiplet
up to terms quadratic in the
field strength  is
\begin{align}
  \label{eq:cpt20action}
  S_g&=\intd{^2x\rd\theta_{0}^+} \bigl(
  -u^0_{ab}\delta^{pq}W_{p-}^a\nabla_{0+}W_{q-}^b
  +u^1_{ab}\delta^{pq}W_{p-}^a\nabla_{1+}W_{q-}^b+i z^p_a W^a_{p-} \bigr)\ ,
\end{align}
where $u^0$ and $u^1$ are the gauge coupling constants which in
general depend on the superfield $\phi$ and similarly for the
 `theta'  terms $z^p$. Both $u^0$ and $u^1$
are not necessarily symmetric in the gauge indices. The above action can be
written in different ways. However there are always field and coupling
constant redefinitions which can bring the action to the above form.

Observe that this action is not an integral over the full $\Xi^{2,0}$
superspace.  Therefore it is not manifestly (2,0)-supersymmetric.
The requirement of invariance under (2,0) supersymmetry imposes the
conditions
\begin{align}
\begin{split}
 J^j{}_i\partial_ju^0 &= -\partial_iu^1 \\
 J^j{}_i\partial_jz^1 &= -\partial_iz^0
 ~.
\end{split}
 \label{eq:cr}
\end{align}
This is most easily seen by verifying that the Lagrangian density is
independent of $\theta_{1}^+$ up to $\theta_{0}^+, x^\plpl,
x^\mimi$-surface terms.  The conditions~\eqref{eq:cr} are the
Cauchy-Riemann equations which imply that $u^0+iu^1$ and $z^1+iz^0$
are {\it holomorphic}.

Indeed provided that the holomorphicity conditions~\eqref{eq:cr} hold,
the action~\eqref{eq:cpt20action}, apart from the theta terms, can be
written as an integral over the $\Xi^{2,0}$ superspace as
\begin{align}
\label{eq:manifest20action}
\begin{split}
  S_g=\intd{^2x\rd\theta_{0}^+\rd\theta_{1}^+} \bigl(&
  \alpha u^0_{ab} W_{0-}^a W_{1-}^b
  + (\alpha-1) u^1_{ab} W_{0-}^a W_{0-}^b \\
&  - \alpha u^1_{ab} W_{1-}^a W_{1-}^b
  + (\alpha-1) u^0_{ab} W_{1-}^a W_{0-}^b
  \bigr)
\end{split}
\end{align}
for any constant $\alpha$. After integrating over the odd coordinate
$\theta_{1}^+$ we recover the action~\eqref{eq:cpt20action}. Observe
that the action~\eqref{eq:manifest20action} simplifies if one takes $u^0, u^1$
to be  symmetric matrices.  In particular one finds that
\begin{align}
  S_g=\intd{^2x\rd\theta_{0}^+\rd\theta_{1}^+} u^0_{ab}W_{0-}^aW_{1-}^b~.
\end{align}

Invariance of the action  \eqref{eq:cpt20action} under gauge transformations
requires that the couplings $u^0$, $u^1$ and $z^p$ satisfy
\begin{align}
\begin{split}
{\cal L}_a u^0_{bc}&=-f^e{}_{ab} u^0_{ec}-f^e{}_{ac} u^0_{ae}
\\
{\cal L}_a u^1_{bc}&=-f^e{}_{ab} u^1_{ec}-f^e{}_{ac} u^1_{ae}
\\
{\cal L}_a z^p_{b}&=-f^c{}_{ab} z^p_c
\ .
\end{split}
\end{align}
The gauge transformations of the gauge multiplet and the sigma model
multiplet that are required to derive the above result are as in
the (1,0)-supersymmetric models studied in the previous sections.

\subsection{The sigma model action}

The part of the action which describes the coupling of the sigma model
(2,0)-multiplet to the gauge multiplet has already been given in
\cite{hgps}.  This action is
\begin{align}
\begin{split}
  S_\sigma =&-i \intd{^2x\rd\theta_{0}^+}\bigl(
  g_{ij}\nabla_{0+}\phi^i\nabla_\mimi\phi^j
  + \nu_{a}W^a_{1-}\bigr)\\
  &-i \intd{^2x\rd t\rd\theta_{0}^+}\bigl(
  H_{ijk}\partial_t\phi^i\nabla_{0+}\phi^j\nabla_{\mimi}\phi^k -
  w_{ia}\partial_t\phi^iW_{0-}^a \bigr)
\end{split}
\end{align}
where $\nu_{a}$ is a function on $M$, possibly locally defined,
given by
\begin{align}
\label{eq:moment10def}
  I^j_{\phantom{j}i} (\xi_{aj} + w_{aj}) &=  -\partial_i \nu_{a}
\end{align}
Under certain conditions the maps $\nu$ can be
thought of as the moment maps of KT geometry
~\cite{ggpp}.

Gauge invariance of the above part of action requires in addition to the
conditions on  $w$, which we have already mentioned in section  \ref{sub:gaugecon},
that $\nu_a$ is
globally defined on $M$ and that
\begin{align}
\mathcal{L}_a \nu_b=-f_{ab}{}^c \nu_c~.
\end{align}

\subsection{The action of the fermionic multiplet}

This part of the action is
\begin{align}
  S_f=\intd{^2x\rd\theta_{0}^+}\bigl(
    h_{AB} \psi_-^A \Nabla_{0+} \psi_-^B +m s_A \psi_-^A
  \big)
\end{align}

Gauge invariance of this part of the action requires the same
conditions as those appearing for the couplings of (1,0)-multiplet
in~\eqref{eq:psigaugecnd}.
 
The conditions required by (2,0)-supersymmetry on the couplings of the
above action are the same as those of the ungauged model and have been
given in~\cite{hw}. These can be easily derived by requiring that the
Lagrangian density is independent from $\theta_{1}^+$ up to $x^\plpl,
x^\mimi, \theta_{0}^+$ surface terms. In particular, we find that
\begin{align}
 \label{eq:susyfcon}
 \begin{split}
 h_{CB} I^C{}_A+h_{CA} I^C{}_B&=0
 \\
 J^j{}_i \nabla_j h_{AB}+\nabla_i h_{AC} I^C{}_B&=0
 \\
J^j{}_i \nabla_j s_A-\nabla_i (s_B I^B{}_A)-{1\over2} \nabla_i
h_{AB} L^B&=0
\\
s_A L^A&={\rm const}\ .
\end{split}
\end{align}
The first condition implies that the fibre metric is hermitian is
respect to the fibre complex structure. It is always possible to
choose such a fibre metric given a fibre complex structure on a bundle
vector bundle $E$.  In the context of sigma models this has been
explained in \cite{phgpb}.  The rest of the conditions can be
considerably simplified if the connection $B$ is chosen such that
$\nabla I=\nabla h=0$. Such connection always exists on a hermitian
vector bundle $E$. In such case, the third equation in
\eqref{eq:susyfcon} implies that $s$ is the real part of a holomorphic
section of ${\cal E}^*$.

\subsection{The action in components}

It is straightforward to write the action $S$ of the
(2,0)-supersymmetric gauge sigma model in components.
In particular we find that the component action of the gauge
multiplet~\eqref{eq:cpt20action} is
\begin{align}
  \begin{split}
    S_g = \intd{^2x}\bigl(& u^0_{ab} F_{\plpl\mimi}^a
 F_{\plpl\mimi}^b+ z^0_a F_{\plpl\mimi}^a
-u^0_{ab} f^a f^b + i z_a^1 f^a
    \\
    &+
    i u^0_{ab} \chi_{0-}^a \nabla_\plpl\chi_{0-}^b +
    i u^0_{ab} \chi_{1-}^a \nabla_\plpl\chi_{1-}^b
     \\
    & -2i u^1_{[ab]} F_{\plpl\mimi}^a f^b
    +2i u^1_{[ab]} \chi_{[0-}^a \nabla_\plpl\chi_{1]-}^b
    \\
    &+i\partial_iz^0_a \lambda_+^i \chi_{0-}^a+
 i\partial_i z^1_a \lambda^i_+  \chi_{1-}
    \\
    & -\lambda^i_+ \partial_iu^0_{ab}(i\chi_{0-}^aF_{\plpl\mimi}^b
    +\chi_{1-}^af^b)  \\
&   +\lambda^i_+\partial_iu^1_{ab}(-\chi_{0-}^af^b
    +i\chi_{1-}^aF_{\plpl\mimi}^b) \bigr)~.
  \end{split}
\end{align}
The component action of the sigma model part is
\begin{align}
  \begin{split}
    S_\sigma = \intd{^2x} \bigl(&
    g_{ij} \nabla_{\plpl}\phi^i \nabla_\mimi\phi^j
    + b_{ij} \partial_\plpl\phi^i \partial_\mimi\phi^j
      +i \lambda_+^i \Nabla^{(+)}_\mimi \lambda_+^j \\
&      -i  \partial_i \lambda_+^i \nu_a \chi_{1-}^a
      -i \nu_a f^a 
      +i g_{ij} \lambda_+^i \chi_{0-}^a \xi_a^{\phantom{a}j}
   +i w_{ia} \lambda_+^i \chi_{0-}^a \\
&    -  A_\plpl^a w_{ia} \partial_\mimi\phi^i
    +  A_\mimi^a w_{ia} \partial_\plpl\phi^i
    -  A_\plpl^a A_\mimi^b \xi_{[b}^{\phantom{[b}i} w_{a]i}
    \bigr)
  \end{split}
\end{align}
and the component action of the fermionic multiplet is
\begin{align}
  \begin{split}
    S_f = \intd{^2x} \bigl(&
    -i h_{AB} \psi_-^A \Nabla_\plpl \psi_-^B
    + h_{AB} \ell^A \ell^B \\
&    - \frac{1}{2} h_{AB} \psi_-^A \psi_-^B \lambda_+^i \lambda_+^j
    G_{ijAB} 
    + m \nabla_i s_A \lambda_+^i \psi_-^A
    + m s_A \ell^A
    \bigr)\ .
  \end{split}
\end{align}
Eliminating the auxiliary fields of the gauge and fermionic
multiplets, we find that
\begin{align}
  \begin{split}
    S = \intd{^2x}\bigl(& 
    g_{ij} \nabla_{\plpl}\phi^i \nabla_\mimi\phi^j
    + b_{ij} \partial_\plpl\phi^i \partial_\mimi\phi^j
    + u^0_{ab} F_{\plpl\mimi}^a F_{\plpl\mimi}^b \\
&   +i u^0_{ab} \chi_{0-}^a \nabla_\plpl\chi_{0-}^b +
    i u^0_{ab} \chi_{1-}^a \nabla_\plpl\chi_{1-}^b
     \\
&     +i \lambda_+^i \tilde\nabla^{(+)}_\mimi \lambda_+^j 
   -i h_{AB} \psi_-^A \Nabla_\plpl \psi_-^B 
   + z^0_a F_{\plpl\mimi}^a \\
&    -\frac{1}{4} m^2 h^{AB} s_A s_B 
     -\frac{1}{4} u_0^{ab} (\nu_a - z^1_a)(\nu_b - z^1_b) \\
&    - \frac{1}{2} h_{AB} \psi_-^A \psi_-^B \lambda_+^i \lambda_+^j
    G_{ijAB} 
    + m \nabla_i s_A \lambda_+^i \psi_-^A \\
&    -  A_\plpl^a w_{ia} \partial_\mimi\phi^i
    +  A_\mimi^a w_{ia} \partial_\plpl\phi^i
    -  A_\plpl^a A_\mimi^b \xi_{[b}^{\phantom{[b}i} w_{a]i} \\
&    -  u_0^{ab} (\nu_a - z^1_a)  u^1_{[cb]} F_{\plpl\mimi}^c 
   - u_0^{ab} u^1_{[ac]} u^1_{[bd]} F_{\plpl\mimi}^c F_{\plpl\mimi}^d \\
&+i\partial_i z^0_a \lambda_+^i \chi_{0-}^a+
 i\partial_i z^1_a \lambda^i_+  \chi_{1-}
    \\
    & -i \partial_i  u^0_{ab} \lambda^i_+ \chi_{0-}^aF_{\plpl\mimi}^b  
   +i \partial_i u^1_{ab} \lambda^i_+\chi_{1-}^aF_{\plpl\mimi}^b \\
&      -i \partial_i \nu_a \lambda_+^i \chi_{1-}^a
      +i g_{ij} \lambda_+^i \chi_{0-}^a \xi_a^{\phantom{a}j}
   +i w_{ia} \lambda_+^i \chi_{0-}^a \\
&  +\frac{1}{2}i u_0^{ab} (\nu_a - z^1_a)(\partial_i u^0_{cb}\chi_{1-}^c
    + \partial_i u^1_{cb}\chi_{0-}^c) \\
&   +\frac{1}{4} u_0^{ab}(\partial_i u^0_{ca}\chi_{1-}^c
      + \partial_i u^1_{cb}\chi_{0-}^c)
    (\partial_i u^0_{db}\chi_{1-}^d + \partial_i u^1_{db}\chi_{0-}^d)\\
&   -i u_0^{ab} u^1_{[ac]} F_{\plpl\mimi}^c 
    (\partial_i u^0_{db}\chi_{1-}^d + \partial_i u^1_{db}\chi_{0-}^d)\\
&     +2i u^1_{[ab]} \chi_{[0-}^a \nabla_\plpl\chi_{1]-}^b
    \bigr)
  \end{split}
\end{align}
where $u_0^{ab}$ is the matrix inverse of $u^0_{(ab)}$,
\begin{align}
  u_0^{ab} u^0_{(bc)} &= \delta^a_{\phantom{a}c}~.
\end{align}
Note that we have assumed that $u^0$ is invertible.

\subsection{Scalar potential and classical vacua}

The scalar potential of the
(2,0)-supersymmetric  gauge theories coupled to sigma model matter is
\begin{align}
V=\frac{1}{4}u_0^{ab}(\nu_a-z^1_a) (\nu_b-z^1_b)+\frac{1}{4}m h^{AB} s_A s_B\ .
\end{align}
The scalar potential in these models is
written as a sum of a `$D$' and an `$F$' term. The classical
supersymmetric vacua of
the theory are those for which
\begin{align}
\nu_a-z^1_a&=0 & s_A&=0\ .
\end{align}
The inequivalent classical vacua are the space of orbits of the gauge group
on the zero set of the section $s$ and  $\nu-z^1$. If the section $s$ and $z^1$
vanish, then the space of inequivalent vacua is the KT reduction $M//G$ of the sigma model
target space $M$. It has been shown in \cite{ggpp} that the space of vacua inherits the KT
structure of the sigma model manifold $M$ and under certain assumptions
is a smooth manifold. However the three-form of the Wess-Zumino term
 on $M//G$ is not necessarily
closed.

\section{$(4,0)$ supersymmetry}

\subsection{The gauge multiplet}

The $(4,0)$ superspace $\Xi^{4,0}$ has coordinates
$(x^\plpl,x^\mimi,\theta_{p}^+)$,  where
$\theta_{p}^+$, $p=0,1,2,3$, are the odd coordinates. The
(4,0)-supersymmetric Yang-Mills multiplet is described by a
connection $A$ in superspace with components
$(A_\plpl,A_\mimi,A_{p+})$, $p=0,1,2,3$. In addition it is required
that these satisfy the supersymmetry constraints \cite{hgps}
\begin{align}\begin{split}
  [ \nabla_{p+},  \nabla_{q+} ] &= 2i\delta_{pq}\nabla_\plpl \\
 [ \nabla_\plpl, \nabla_\mimi ] &= F_{\plpl\mimi} \\
 [ \nabla_{p+},  \nabla_\mimi ] &= W_{p-} \\
\nabla_{p+} W_{q-}&=\frac{1}{2} \epsilon_{pq}{}^{p'q'}\nabla_{p'+} W_{q'-} \ ,
\end{split}\end{align}
where $p,q,p'q'=0,\ldots,3$ and $p\neq q$ in the last condition. (We have suppressed
gauge indices.)
Jacobi identities imply that
\begin{align}
  \nabla_{p+} W_{q-} + \nabla_{q+} W_{p-}
&=  2i\delta_{pq}F_{\plpl\mimi}
\end{align}

The components of the gauge multiplet are
\begin{gather}
  \chi_{p-} = W_{p-}|
\mspace{100mu} i F_{\plpl\mimi} = \nabla_{0+}W_{0-}| 
 \nonumber\\
  f_r = \nabla_{0+}W_{r-}| \quad (r=1,2,3)
\end{gather}
The first four fields, $(\chi_{p-}: p=0,1,2,3)$, are the gaugini which are real
chiral fermions in two dimensions, $F_{\plpl\mimi}$ is the field
strength and $\{f_r: r=1,2,3\}$ are the auxiliary fields.

\subsection{The sigma model multiplet}

Let $M$ be a hyper-K\"ahler manifold with torsion (HKT).
This implies that $M$ admits  a  hypercomplex structure $\{J_r: r=1,2,3\}$
the metric $g$ on $M$ is tri-hermitian and the hypercomplex structure is
parallel with respect to a metric connection with torsion the three-form $H$, $\nabla^{(+)}J_r=0$;
(see \cite{phgpt, ggks} for more details).
In addition we assume that the gauge group $G$ acts on $M$ preserving
the metric, three-form $H$ and the hypercomplex structure. The latter condition
implies that
\begin{align}
  \mathcal{L}_a J_r=0 \ ,
\end{align}
where the Lie derivative $\mathcal{L}_a$ 
is along vector field $\xi_a$ generated by the
action of $G$ on $M$.  The sigma model fields are
maps $\phi:\Xi^{4,0}\rightarrow M$ into the HKT manifold $M$ which
in addition satisfy
\begin{align}
\nabla_{r+}\phi^i= J_r{}^i{}_j \nabla_{0+}\phi^j \ ,
\end{align}
where $\nabla_{p+}\phi=D_{p+}\phi^i+ A_{p+}^a \xi_a^i$ and $r=1,2,3$.
We remark that the algebra of (4,0) supersymmetry transformations
closes as a consequence of the HKT condition we imposed on $M$.

The components of the sigma model multiplet $\phi$ are as follows,
\begin{align}
\phi^i&=\phi^i| & \lambda^i_{+}=&\nabla_{0+}\phi^i|~.
\end{align}

\subsection{The fermionic multiplet}

Let $E$ be a vector bundle over $M$ equipped with a connection $B$ and a
fibre (almost) hypercomplex structure $\{I_r: r=1,2,3\}$.
The fermionic multiplet $\psi_-$ is a section of $\phi^*E\otimes S_-$
over the $\Xi^{4,0}$, where $S_-$ is a spin bundle over $\Xi^{4,0}$.
In addition the fermionic multiplet $\psi_-$ satisfies
\begin{align}
  \Nabla_{r+}\psi_-^A= I_r{}^A{}_B \Nabla_{0+}\psi_-^B+ \frac{1}{2}m L_r^A
  \label{con40con}
\end{align}
where $\{L_r: r=1,2,3\}$ are sections of $E$
and
\begin{align}
  \Nabla_{p+}\psi_-^A=
  D_{p+}\psi^A+\nabla_{p+}\phi^i B_i{}^A{}_B \psi_-^B+ A^a_{p+}
  U_a{}^A{}_B  ~,
\end{align}
$p=0,1,2,3$.

Compatibility of the  condition \eqref{con40con} with gauge transformations
requires that
\begin{align}
\begin{split}
{\cal L}_a I_r^A{}_B&= U_a{}^A{}_C I_r^C{}_B-I_r^A{}_C U_a{}^C{}_B
\\
{\cal L}_a L_r^A&=  U_a{}^A{}_B L_r^B\ .
\end{split}
\end{align}
These are the conditions for the gauge transformations and the (2,0)-supersymmetry
transformations to commute.

The conditions required for the closure of the (4,0) supersymmetry
algebra are similar to those found for the ungauged (4,0) model
in \cite{hgpt,phgp}. Here to simplify the analysis, we shall in addition assume that
the fibre hypercomplex structure is parallel with respect to $\nabla$, ie $\nabla I_r=0$.
(See \cite{phgpb} for a discussion on the conditions required for the existence
of such a connection $\nabla$ on the vector bundle $E$.) The more general case
can be easily derived but we shall not use these results later.
In this special case, we find that
\begin{align}\begin{split}
G_{kl}{}^A{}_B J_{r}{}^k{}_i
 J_{s}{}^l{}_j+G_{kl}{}^A{}_B J_{s}{}^k{}_i J_{r}{}^l{}_j
=2\delta_{rs}G_{ij}{}^A{}_B
\\
J_r{}^J{}_i \nabla_j L_s^A+ J_s{}^J{}_i \nabla_j L_r^A- I_r{}^A{}_B
\nabla_i L_s^B-I_s{}^A{}_B \nabla_i L_r^B =0\ .
\end{split}\end{align}
The first condition implies that the curvature $G$ of the vector bundle $E$ is
an (1,1)-form with respect to all three complex 
structures $J_r$. Observe that the
diagonal relation $r=s$ implies all the rest.
The diagonal part of the second condition implies that each section $L_r$ is a
holomorphic section with respect to the pair $(J_r, I_r)$.

The components of the fermionic multiplet are as follows:
\begin{align}
\psi_-^A&=\psi_-^A| & \ell^A=&\Nabla_{0+}\psi_-^A| \ ,
\end{align}
where $\psi_-$ is a two-dimensional real chiral fermion and $\ell$ is
an auxiliary field.

\subsection{Action}

The action of the (4,0)-supersymmetric gauged sigma model can be written
as
\begin{align}
  S=S_g+S_\sigma+S_f  \ ,
\end{align}
where $S_g$ is the action of the gauge multiplet, $S_\sigma$ is the
action of the sigma-model multiplet and $S_f$ is the action of the
fermionic multiplet. We shall describe each term separately.

\subsection{The gauge multiplet action}

The most general action for the (4,0)-supersymmetric gauge multiplet
up to terms quadratic in the
field strength  is
\begin{align}
  \label{eq:cpt40action}
  S_g&=\intd{^2x\rd\theta_{0}^+} \bigl(
  -u^0_{ab}\delta^{pq}W_{p-}^a\nabla_{0+}W_{q-}^b
  +\sum^3_{r=1}u^r_{ab}\delta^{pq}W_{p-}^a\nabla_{r+}W_{q-}^b+i z^p_a W^a_{-p} \bigr)
\end{align}
where $\{u^p: p=0,1,2,3\}$ and $\{z^p: p=0,1,2,3\}$ are the gauge coupling constants which in
general depend on the superfield $\phi$ and $u^p$ are not necessarily symmetric
in the gauge indices. The above action can be
written in different ways. However there are always field and coupling
constant redefinitions which bring the action to the above form.

Observe that this action is not an integral over the full $\Xi^{4,0}$
superspace.  Therefore it is not manifestly (4,0)-supersymmetry.
Define $J_{0\phantom{i}j}^{\phantom{0}i}=\delta^i_{\phantom{i}j}$. The
requirement of invariance the action 
\eqref{eq:cpt40action}  under (4,0) supersymmetry imposes the
conditions
\begin{align}\begin{split}
  \label{eq:cr40}
      J_{p\phantom{j}i}^{\phantom{p}j} \partial_j u_q
  &= \frac{1}{2} \epsilon_{pq}{}^{p'q'}
     J_{p'}{}^j{}_i \partial_ju_{q'}
  \quad (p\neq q)
   \\
  \partial_i u_0
    = J_{1\phantom{j}i}^{\phantom{1}j} \partial_j u_1
    &= J_{2\phantom{j}i}^{\phantom{2}j} \partial_j u_2
    = J_{3\phantom{j}i}^{\phantom{3}j} \partial_j u_3
  ~,
\end{split}\end{align}
and
$\{u^r: r=1,2,3\}$ are {\it symmetric} in the gauge indices. In addition,
we have
\begin{align}
\label{eq:sus40}
\begin{split}
J_r{}^j{}_i \partial_jz^r&=-\partial_i z^0\\
J_p{}^j{}_i \partial_j z_q&=-{1\over2}
 \epsilon_{pq}{}^{p'q'} J_{p'}{}^k{}_i \partial_k z_{q'}\ .
\end{split}
\end{align}
The conditions \eqref{eq:cr40} and \eqref{eq:sus40} imply that in fact
$\{u^p: p=0,\dots, 3\}$ and $\{z^p: p=0,\dots, 3\}$ are constant, i.e.\ 
independent from the sigma model superfield $\phi$.  The above
conditions are most easily derived by verifying that the Lagrangian
density is independent of $\theta_{r}^+$ up to surface terms in
$x^\plpl$, $x^\mimi$ and $\theta_{0}^+$.  In addition gauge invariance
requires that the coupling constants $u^p$ and $z^p$ satisfy the
condition
\begin{align}\begin{split}
f^d{}_{ab} u^p_{dc}+ f^d{}_{ac} u^p_{bd}&=0
\\
f^c{}_{ab} z^p_{c}&=0\ .
\end{split}\end{align}
In particular $u^r$ should be proportional to an invariant quadratic
form on the Lie algebra of the gauge group $G$. If $G$ semi-simple,
the condition on $z^p$ implies that $z^p=0$. If $G$ is abelian, then
the above conditions are satisfied for any constants $u^p$ and $z^p$.

\subsection{The sigma model multiplet action}

This part of the action has already been described in \cite{hgps}.
Here we shall summarise some of results relevant to this paper.
The action of this multiplet is
\begin{align}
\begin{split}
  S_\sigma =&-i \intd{^2x\rd\theta_{0}^+}\bigl(
    g_{ij} \nabla_{0+}\phi^i \nabla_\mimi\phi^j
    +\sum_{r=1}^3 \nu_{ra} W_{r-}^a
  \bigr)\\
  &-i \intd{^2x\rd t\rd\theta_{0}^+}\bigl(
  H_{ijk}\partial_t\phi^i\nabla_{0+}\phi^j\nabla_{\mimi}\phi^k -
  w_{ia}\partial_t\phi^iW_{0-}^a \bigr)\ ,
\end{split}
\end{align}
where $\nu_{a}$ is a function on $M$, possibly locally defined,
given by
\begin{align}
\label{eq:momentp0def}
  I_{r\phantom{j}i}^{\phantom{r}j} (\xi_{aj} + w_{aj}) 
&=  -\partial_i\nu_{ra}\ .
\end{align}
It has been shown in~\cite{ggpp} that under certain
conditions $\nu$ is a moment map of HKT geometry.

The gauge transformations of the superfield $\phi$ are
\begin{align}
\delta\phi^i &= \lambda^a\xi_a^{\phantom{a}i}(\phi)
\end{align}
Gauge invariance of the above action requires that  $w$
should satisfy the conditions  mentioned in section \ref{sub:gaugecon},
the moment maps should 
be globally defined on the sigma model target space $M$ and
\begin{gather}
  \mathcal{L}_a \nu_{ra} = -f_{ab}^{\phantom{ab}c} \nu_{rc} ~.
\end{gather}

\subsection{The action of the fermionic multiplet}

This part of the action is
\begin{align}
  S_f=\intd{^2x\rd\theta_{0}^+}\bigl(
    h_{AB} \psi_-^A \Nabla_{0+} \psi_-^B +m s_A \psi_-^A
  \big)
\end{align}

Gauge invariance of this part of the action requires the same
conditions as those appearing for the couplings of (1,0)-multiplet
 in~\eqref{eq:psigaugecnd}.

The conditions required by (4,0)-supersymmetry on the couplings of the
above action are the same as those of the ungauged model and have been
given in~\cite{phgp}. These can be easily derived by requiring that
the Lagrangian density is independent from $\theta_{r}^+$ up to
 $x^\plpl, x^\mimi, \theta_{0}^+$ surface terms. In particular, we find that
 \begin{align}
 \label{eq:susyf40con}
 \begin{split}
 h_{CB} I_r{}^C{}_A+h_{CA} I_r{}^C{}_B&=0
 \\
 J_r{}^j{}_i \nabla_j h_{AB}+\nabla_i h_{AC} I_r{}^C{}_B&=0
 \\
J_r{}^j{}_i \nabla_j s_A-\nabla_i s_B I_r{}^B{}_A-{1\over2} \nabla_i
h_{AB} L_r^B&=0
\\
s_A L_r^A={\rm const}\ .
\end{split}
\end{align}
To derive the above conditions we have used that $\nabla I_r=0$ as we
have assumed in the construction of the fermionic multiplet. The first
condition implies that the fibre metric is tri-hermitian.  These
conditions can be further simplified if the connection $B$ is chosen
such that $\nabla h=0$.  In such case, the third equation in
(\ref{eq:susyf40con}) implies that $s$ is the real part of three
holomorphic sections of ${\cal E}^*$ each with respect to the three
doublets $(J_r, I_r)$ of complex structures, ie $s$ is triholomorphic.

\subsection{The action in components and scalar potential}

The part of the action of the theory involving the
kinetic term of the gauge multiplets \eqref{eq:cpt40action} can be easily expanded in components
as follows:
\begin{align}
  \begin{split}
    S_g = \intd{^2x} \bigl( & u^0_{ab} F_{\plpl\mimi}^a F_{\plpl\mimi}^b
      +i u^0_{ab} \delta^{pq} \chi_{p-}^a \nabla_{\plpl}\chi_{q-}^b\\
&   + z^0_a F^a_{\plpl\mimi}
    - \sum_r (u^0_{ab} f_r^a f_r^b+i z^r_a f_r^a )\\
&   + 2i u^1_{ab} \chi_{[0-}^a \nabla_\plpl\chi_{1]-}^b
    + 2i u^1_{ab} \chi_{[2-}^a \nabla_\plpl\chi_{3]-}^b \\
&   + 2i u^2_{ab} \chi_{[0-}^a \nabla_\plpl\chi_{2]-}^b
    + 2i u^2_{ab} \chi_{[1-}^a \nabla_\plpl\chi_{3]-}^b \\
&   + 2i u^3_{ab} \chi_{[0-}^a \nabla_\plpl\chi_{3]-}^b
    + 2i u^3_{ab} \chi_{[1-}^a \nabla_\plpl\chi_{2]-}^b     
    \bigr) \ .
  \end{split}
\end{align}
To derive this we have used that $\{u^p:0,\dots, 3\}$ are constant and $\{u^r:1,2,3\}$ symmetric
in the gauge indices.

The part of the action that contains the kinetic term and the
Wess-Zumino term of the sigma model fields in components is as
follows,
\begin{align}
  \begin{split}
    S_\sigma = \intd{^2x} \bigl(&
    g_{ij} \nabla_{\plpl}\phi^i \nabla_\mimi\phi^j
    + b_{ij} \partial_\plpl\phi^i \partial_\mimi\phi^j
      +i \lambda_+^i \tilde\nabla^{(+)}_\mimi \lambda_+^j \\
&      -i \sum_r ( \nu_{ra} f_r^a 
+ \partial_i \nu_{ra} \lambda_+^i \chi_{r-}^a)\\
&      +i g_{ij} \lambda_+^i \chi_{0-}^a \xi_a^{\phantom{a}j}
   +i w_{ia} \lambda_+^i \chi_{0-}^a \\
&   -  A_\plpl^a w_{ia} \partial_\mimi\phi^i
    +  A_\mimi^a w_{ia} \partial_\plpl\phi^i
    -  A_\plpl^a A_\mimi^b \xi_{[b}^{\phantom{[b}i} w_{a]i}
    \bigr)
  \end{split}
\end{align}
The part of the action that contains
 the kinetic term the fermionic multiplet in components is as follows:
\begin{align}
  \begin{split}
    S_f = \intd{^2x} \bigl(&
    -i h_{AB} \psi_-^A \Nabla_\plpl \psi_-^B
    + h_{AB} \ell^A \ell^B 
    - \frac{1}{2} h_{AB} \psi_-^A \psi_-^B \lambda_+^i \lambda_+^j
    G_{ijAB} \\
&    + m \nabla_i s_A \lambda_+^i \psi_-^A
    + m s_A \ell^A
    \bigr)
  \end{split}
\end{align}
After eliminating the auxiliary fields of both the gauge multiplet and
the fermionic multiplet, we find that the action of
(4,0)-supersymmetric gauge theories coupled to sigma models is
\begin{align}
  \begin{split}
    S_g+S_f = \intd{^2x}\bigl(& 
    u^0_{ab} F_{\plpl\mimi}^a F_{\plpl\mimi}^b 
   -i h_{AB} \psi_-^A \Nabla_\plpl \psi_-^B \\
&  +i \delta^{pq} u^0_{ab} \chi_{p-}^a \nabla_\plpl\chi_{q-}^b 
     \\
&   + 2i u^1_{ab} \chi_{[0-}^a \nabla_\plpl\chi_{1]-}^b
    + 2i u^1_{ab} \chi_{[2-}^a \nabla_\plpl\chi_{3]-}^b \\
&   + 2i u^2_{ab} \chi_{[0-}^a \nabla_\plpl\chi_{2]-}^b
    + 2i u^2_{ab} \chi_{[1-}^a \nabla_\plpl\chi_{3]-}^b \\
&   + 2i u^3_{ab} \chi_{[0-}^a \nabla_\plpl\chi_{3]-}^b
    + 2i u^3_{ab} \chi_{[1-}^a \nabla_\plpl\chi_{2]-}^b  \\
&   - \frac{1}{2} h_{AB} \psi_-^A \psi_-^B \lambda_+^i \lambda_+^j
    G_{ijAB} 
    + m \nabla_i s_A \lambda_+^i \psi_-^A \\
&     - \frac{1}{4} u_0^{ab} \sum_r (\nu_{ra} - z^r_a)(\nu_{rb} - z^r_b) 
    -\frac{1}{4} m^2 h^{AB} s_A s_B \\
&   + z^0_a F_{\plpl\mimi}^a 
      + \sum_p \partial_i z^p_a \lambda_+^i \chi_{r-}^a
    \bigr)
  \end{split}
\end{align}
It is straightforward to write the action $S$ of the
(2,0)-supersymmetric gauge sigma model in components. In particular
 we find the following:
\begin{align}
  \begin{split}
    S = \intd{^2x}\bigl(& 
    g_{ij} \nabla_{\plpl}\phi^i \nabla_\mimi\phi^j
    + b_{ij} \partial_\plpl\phi^i \partial_\mimi\phi^j
    + u^0_{ab} F_{\plpl\mimi}^a F_{\plpl\mimi}^b \\
&   +i u^0_{ab} \delta^{pq} \chi_{p-}^a \nabla_\plpl\chi_{q-}^b 
     \\
&     +i \lambda_+^i \tilde\nabla^{(+)}_\mimi \lambda_+^j 
   -i h_{AB} \psi_-^A \Nabla_\plpl \psi_-^B 
   + z^0_a F_{\plpl\mimi}^a \\
&    -\frac{1}{4} m^2 h^{AB} s_A s_B 
     - \frac{1}{4} u_0^{ab} \sum_r (\nu_{ra} - z^r_a)(\nu_{rb} - z^r_b) \\
&    - \frac{1}{2} h_{AB} \psi_-^A \psi_-^B \lambda_+^i \lambda_+^j
    G_{ijAB} 
    + m \nabla_i s_A \lambda_+^i \psi_-^A \\
&    -  A_\plpl^a w_{ia} \partial_\mimi\phi^i
    +  A_\mimi^a w_{ia} \partial_\plpl\phi^i
    -  A_\plpl^a A_\mimi^b \xi_{[b}^{\phantom{[b}i} w_{a]i} \\
&+i\partial_i z^0_a \lambda_+^i \chi_{0-}^a
    \\
&          +i g_{ij} \lambda_+^i \chi_{0-}^a \xi_a^{\phantom{a}j}
   +i w_{ia} \lambda_+^i \chi_{0-}^a 
  + i\sum_r \partial_i (z^r_a-\nu_{ra}) \lambda_+^i \chi_{r-}^a
    \bigr)
  \end{split}
\end{align}

\subsection{Scalar Potential and Classical Vacua}

The scalar potential of the (4,0)-supersymmetric gauged sigma models
is
\begin{align}
V=\frac{1}{4}\sum^3_{r=1}u_0^{ab}\nu_{ra} \nu_{rb}+\frac{1}{4}m h^{AB} s_A s_B\ ,
\end{align}
where we have absorbed the constants $z^r_a$ into the definition
of the moment maps $\nu_{ra}$.
The scalar potential in these models is
written as a sum of a `$D$' and an `$F$' term. The classical
supersymmetric vacua of
the theory are those for which
\begin{align}
\nu_{ra}&=0 & s_A&=0\ .
\end{align}
The inequivalent classical vacua are the space of orbits of the gauge group
on the zero set of the section $s$ and the HKT moment maps $\nu_r$. If the section $s$
vanishes, then the space of inequivalent vacua is  the theory is  the HKT reduction $M//G$ of the sigma model
target space $M$. It has been shown in \cite{ggpp} that under certain assumptions the space of vacua inherits the HKT
structure of the sigma model manifold $M$ and it is a smooth space.
However the three-form of the Wess-Zumino term on $M//G$ is not necessarily
closed.

\section{$(1,1)$ supersymmetry}

\subsection{The gauge multiplet}

The $(1,1)$ superspace $\Xi^{1,1}$ has coordinates
$(x^\plpl,x^\mimi,\theta^+,\theta^-)$, where $\theta^{\pm}$ are
Grassman valued odd coordinates. The (1,1)-supersymmetric Yang-Mills
multiplet is described by a connection $A$ in superspace with
components $(A_\plpl,A_\mimi,A_+,A_-)$. In addition it is required
that these satisfy the supersymmetry constraints \cite{hgps}
\begin{align}
  [ \nabla_+, \nabla_- ] &= W & [ \nabla_\plpl, \nabla_\mimi ] &=
  F_{\plpl\mimi}  \nonumber\\
  [ \nabla_+, \nabla_+ ] &= 2i\nabla_\plpl & [ \nabla_-, \nabla_- ] &=
  2i\nabla_\mimi
\end{align}
(We have suppressed the gauge indices.)
The Jacobi identities imply that
\begin{gather}
  [\nabla_+, \nabla_\mimi ] = i\nabla_-W
\mspace{100mu} [\nabla_-, \nabla_\plpl ] = i\nabla_+W
 \nonumber\\
  F_{\plpl\mimi} = \nabla_+ \nabla_-W
\end{gather}
It is worth mentioning that the (1,1)-supersymmetric gauge multiplet can be
constructed from a scalar superfield. This allows for the possibility of
non-polynomial couplings between the sigma model
 multiplet and the gauge multiplet.
The components of the gauge multiplet are
\begin{align}
  W &= W|
& F_{\plpl\mimi} &= \nabla_+\nabla_-W|\nonumber \\
  \chi_+ &= \nabla_+ W|
& \chi_- &= \nabla_- W|
\end{align}
The field, $W$, is a scalar, $\chi_+$,$\chi_-$ are
gaugini which are real chiral fermions in two dimensions, and
$F_{\plpl\mimi}$ is the (gauge) field strength.

\subsection{The sigma model multiplet}

Let $M$ be a Riemannian manifold with metric $g$ and locally defined
two-form $b$.  We take the gauge group $G$ to act on $M$
with isometries preserving the Wess-Zumino three-form $H=\rd b$
and generating the vector fields $\xi_a$ as in section \ref{sub:data}. In addition
we assume that the one-form $w$ satisfies the conditions of section \ref{sub:gaugecon}.

The  sigma model (1,1) multiplet $\phi$ is a map from the (1,1) superspace $\Xi^{1,1}$ into the
Riemannian manifold $M$.
The components of $\phi$ are as follows,
\begin{align}
  \phi^i &= \phi^i|
& \ell^i &= \nabla_+\nabla_-\phi^i| \nonumber\\
  \lambda^i_{+}&=\nabla_{+}\phi^i|
& \lambda^i_{-}&=\nabla_{-}\phi^i|
 ~,
\end{align}
where $\nabla_+\phi^i=D_+\phi^i+ A_+^a\xi_a^i$ and similarly for $\nabla_-$.
Observe that the first two components of $(1,1)$ superfield $\phi$
can be identified with the two components of  a $(1,0)$ superfield $\phi$
while the latter two components can be identified with those of  a $(1,0)$
fermionic superfield $\psi_-$. The vector bundle associated with this
fermionic multiplet is the tangent bundle of $M$.

\subsection{Action}

The action of the (1,1)-supersymmetric gauged sigma model can be written
as sum of three terms,
\begin{align}
  S &= S_g + S_\sigma +S_p  \ ,
\end{align}
where $S_g$ is the action of the gauge multiplet, $S_\sigma$ is the
action of the sigma-model multiplet and $S_p$ is a potential term.
We shall describe each term separately.

\subsection{The gauge multiplet action}

An action of the (1,1)-supersymmetric gauge multiplet is most easily
written in (1,1) superspace.
In particular we have
\begin{align}
  S_g = \intd{^2x\rd\theta^+\rd\theta^-}\bigl(
    - u_{ab} \nabla_+W^a \nabla_- W^b
    + \frac{1}{2} v_{ab} W^a W^b+ z_a W^a
    \bigr)\ ,
    \label{f11act}
\end{align}
where $u_{ab}=u_{ab}(\phi)$, $u$ is not necessarily symmetric in the gauge indices,
$v_{ab}=v_{ab}(\phi)$ and the theta term $z_a=z_a(\phi)$. Of course this action is manifestly (1,1)-supersymmetric
because it is an integral over full superspace. Gauge invariance imposes the additional
conditions
\begin{align}
\begin{split}
{\cal L}_a u_{bc}&=-f^d{}_{ab} u_{dc}-f^d{}_{ac} u_{bd}
\\
{\cal L}_a v_{bc}&=-f^d{}_{ab} v_{dc}-f^d{}_{ac} v_{bd}
\\
{\cal L}_a z_b&=-f^d{}_{ab} z_d
\end{split}
\end{align}
on the couplings $u$,$v$ and $z$.

The action~\eqref{f11act} can be easily expanded in components
to find
\begin{align}
  \begin{split}
    S_g = \intd{^2x} \bigl(& 
    + u_{ab} F_{\plpl\mimi}^a F_{\plpl\mimi}^b
     - u_{ab} \nabla_\plpl W^a \nabla_\mimi W^b \\
&   + i u_{ab} \nabla_\plpl\chi_-^a \chi_-^b
      - i u_{ab} \chi_+^a \nabla_\mimi\chi_+^b \\
&      + \partial_i u_{ab} F_{\plpl\mimi}^a \lambda_+^i \chi_-^b
       - \partial_i u_{ab} \lambda_-^i \chi_+^a F_{\plpl\mimi}^b \\
&       + i \partial_i u_{ab} \lambda_+^i \chi_+^a \nabla_\mimi W^b 
      + i \partial_i u_{ab} \lambda_-^i \nabla_\plpl W_a \chi_-^b\\
&        + u_{ab} f^b_{\phantom{b}cd} \chi_+^a \chi_-^c W^d 
      - \nabla_i\partial_j u_{ab} \lambda_+^i \lambda_-^j \chi_+^a \chi_-^b
       - \partial_i u_{ab} \ell^i \chi_+^a \chi_-^b \\
&      + v_{ab} \chi_+^a \chi_-^b  
      + v_{ab} W^a F_{\plpl\mimi}^b \\
&     + \partial_i v_{ab} \lambda_+^i W_a \chi_-^b
      - \partial_i v_{ab} \lambda_-^i \chi_+^a W^b \\
&      + \frac{1}{2} \nabla_i\partial_j v_{ab} \lambda_+^i
         \lambda_-^j W^a W^b
       + \frac{1}{2} \partial_i v_{ab} \ell^i W^a W^b \\
&      + z_a F^a_{\plpl\mimi}
       +\partial_i z_a \lambda_+^i \chi_-
       -\partial_i z_a \lambda_-^i \chi_+\\
&     +  \ell^i\partial_i z_a W^a
      +\lambda_+^i \lambda_-^j \nabla_i\partial_j z_a W^a
    \bigr)\ .
  \end{split}
\end{align}

\subsection{The sigma model multiplet action and potential term}

A (1,1)-supersymmetric gauged
 sigma model action has been given in  \cite{hgps}.
This action can be written as
\begin{align}
\begin{split}
  S_\sigma =& \intd{^2x\rd\theta^+\rd\theta^-}
    g_{ij} \nabla_+\phi^i \nabla_-\phi^j \\
  &+ \intd{^2x\rd t\rd\theta^+\rd\theta^-}\bigl(
  H_{ijk} \partial_t\phi^i \nabla_+\phi^j \nabla_-\phi^k -
  w_{ia} \partial_t\phi^i W_-^a
  \bigr)
\end{split}
\end{align}
This can be rewritten without the $t$ integration as
\begin{align}
\begin{split}
  S_\sigma = \intd{^2x\rd\theta^+\rd\theta^-}\bigl(&
    g_{ij} \nabla_+\phi^i \nabla_-\phi^j
    + b_{ij} D_+\phi^i D_-\phi^j \\
&   - A_+^a w_{ia} D_-\phi^i
    - A_-^a w_{ia} D_+\phi^j
    + A_-^a A_+^b \xi_{[b}^i w_{a]i}
  \bigr)\ .
\end{split}
\end{align}

It is straightforward to add a potential term to the above actions as
\begin{align}
S_p=\intd{^2x\rd\theta^+\rd\theta^-} h
\end{align}
where $h=h(\phi)$ is a function of the superfield $\phi$.
Gauge invariance of the above action requires that $w$ should satisfy the
conditions stated in section \ref{sub:gaugecon}.

\subsection{A generalisation of the action}

The action of (1,1)-supersymmetric gauge theory presented above can be
generalized by allowing the various couplings of the theory to depend
on the scalar component of the gauge multiplet superfield. Supersymmetry
then requires  additional fermionic couplings. The
new theory can be organised as a (1,1)-supersymmetric sigma model
which has target space $L=M\times {\cal L}(G)$, where ${\cal L}(G)$
is the Lie algebra of the gauge group $G$. The various allowed
couplings are restricted by two-dimensional Lorentz invariance,
supersymmetry and gauge invariance. The superfields of
the (1,1)-supersymmetric gauge theory coupled to sigma model
matter are maps $Z=(\phi, W)$ from the (1,1) superspace $\Xi^{1,1}$ into
$L$, where $\phi$ is the usual (1,1) sigma model superfield and $W$
is the (1,1) gauge theory multiplet.
We again allow the gauge group  $G$ to act on $L$ with a group action
on $M$ and the adjoint action on ${\cal L}(G)$.
The vector fields generated by such a group action are
\begin{align}
\xi_a=\xi_a^A\partial_A= \xi_a^i\partial_i+ W^b f^c{}_{ab} \partial_c
\label{nkill}
\end{align}
where $A=(i,a)$, the component $\xi^i$ is allowed to depend on both $\phi$ and $W$, and
the partial derivative with the gauge index denotes differential with
respect to $W$.

Next we introduce a metric $g$ and a Wess-Zumino term $H$ on $L$ and assume
that the gauge group $G$ acts on $L$ with isometries leaving the Wess-Zumino
term $H$ invariant. We also define $w$ as $i_{\xi_{a}}H=dw_a$, where $\xi_a$
is the new Killing vector field \eqref{nkill}.
Then an action can be written for this new sigma model as
\begin{align}
\begin{split}
  S_\sigma =& \intd{^2x\rd\theta^+\rd\theta^-}
   \big( g_{AB} \nabla_+Z^A \nabla_-Z^B+ h \big)\\
  &+ \intd{^2x\rd t\rd\theta^+\rd\theta^-}\bigl(
  H_{ABC} \partial_tZ^A \nabla_+Z^B \nabla_-Z^C -
  w_{Ba} \partial_tZ^B W_-^a
  \bigr)
\end{split}
\end{align}
where $h$ is a function which depends on $Z$.
This action is clearly supersymmetric
 because it is a full (1,1) superspace integral.
Gauge invariance requires that $w$ above satisfies all the conditions stated
in section \ref{sub:gaugecon} but for the group action 
with associated vector fields \eqref{nkill} and a Wess-Zumino term in $L$.
In addition, gauge invariance requires that
\begin{align}
{\cal L}_a h=0
\end{align}
where the Lie derivative is with respect to the vector field \eqref{nkill}.

\subsection{Scalar Potential}

To compute the scalar potential we express the action in components and
eliminate the auxiliary field of the sigma model superfield $\phi$
from the action using the field equations. The scalar potential is
\begin{align}
V(W,\phi)=\frac{1}{4} g^{ij} \partial_i h \partial_j h~,
\end{align}
where $g^{ij}$ is the inverse of the restriction of the metric of $L$ on $M$.
Observe that $V$ depends on both the sigma model scalar $\phi$ and
the gauge multiplet scalar $W$.
 The classical supersymmetric vacua
of the theory are those values of $(\phi,W)$ for which
$\partial_i h=0$. For example, for the special (1,1)-supersymmetric
model investigated in the beginning of the section,
$V=\frac{1}{4} g^{ij} (\partial_i h
+\partial_i z_a W^a) (\partial_j h+\partial_j z_b W^b)$,
where in this case $h=h(\phi)$.

\section{$(2,1)$ supersymmetry}

\subsection{The gauge multiplet}

The $(2,1)$ superspace $\Xi^{2,1}$ has coordinates
$(x^\plpl,x^\mimi,\theta_{p}^+,\theta^-)$, 
where $(x^\plpl,x^\mimi)$ are the even and
$(\theta_{0}^+$,
$\theta_{1}^+$, $\theta^{-})$ are odd coordinates. The
(2,1)-supersymmetric Yang-Mills multiplet is described by a connection
$A$ in superspace with components $(A_\plpl,A_\mimi,A_{p+},A_-)$,
$p=0,1$. In addition it is required that these satisfy the
supersymmetry constraints \cite{hgps}
\begin{align}
  [ \nabla_{p+}, \nabla_- ] &= W_p
& [ \nabla_\plpl, \nabla_\mimi ] &=  F_{\plpl\mimi}
\nonumber\\
  [ \nabla_{p+}, \nabla_{q+} ] &= 2i \delta_{pq} \nabla_\plpl
& [ \nabla_-, \nabla_- ] &=  2i\nabla_\mimi \ .
\end{align}
We have suppressed the gauge indices. We remark that $W_p$ are scalar
superfields.
The Jacobi identities imply that
\begin{gather*}
\begin{align}
[\nabla_{p+}, \nabla_\mimi ] &= i\nabla_-W_p
& [\nabla_-, \nabla_\plpl ] &= i\nabla_{0+} W_0
 \nonumber\\
  F_{\plpl\mimi}^a &= \nabla_{0+} \nabla_- W_0^a 
& \nabla_{1+}W_1&=\nabla_{0+}W_0
\end{align}\label{jac21c}  
\\
  \nabla_{1+}W_0+\nabla_{0+}W_1=0\ . 
\end{gather*}
The two scalar superfields $(W_0, W_1)$ can be viewed as a map $W$
from the (2,1) superspace $\Xi^{2,1}$ into ${\cal L}G\otimes \bR^2$,
where ${\cal L}G$ is the Lie algebra of the group $G$. Next introduce
a complex structure $I={\rm Id}\otimes \epsilon$ in ${\cal L}G\otimes \bR^2$ 
where $\epsilon$ is the constant complex structure in $\bR^2$
with $\epsilon^0{}_1=-1$.  The last two conditions in \eqref{jac21c}
can be expressed as
\begin{align}
\nabla_{1+} W^{ap}=\epsilon^p{}_q \nabla_{0+}W^{aq}\ .
\label{g21con}
\end{align}
In fact this implies that $W$ is a 
covariantly chiral superfield, $(\nabla_{1+}+i\nabla_{0+}) (W_1+iW_0)=0$.

The components of the gauge superfields $W_p$ are
\begin{align}
  W_p &= W_p| 
& F_{\plpl\mimi} &= \nabla_{0+}\nabla_-W_0|
& f &= \nabla_{0+}\nabla_- W_1| \nonumber\\
  \chi_{0+} &= \nabla_{0+} W_0|
& \chi_{1+} &= \nabla_{0+} W_1 |
& \chi_{p-} &= \nabla_- W_p|
\ ,
\end{align}
where $W_p$ are scalars,  $\chi_{p+}$ and $\chi_{p-}$ 
are the gaugini which are real chiral fermions in two dimensions,
$F_{\plpl\mimi}$ is the field strength and $f$ is a real auxiliary  field.
As in the (1,1)-supersymmetric gauge theory, the 
gauge multiplet is determined by scalar superfields. This
will lead again to non-polynomial
 interactions between the gauge and sigma multiplets of the theory.

\subsection{The sigma model multiplet}

Let $M$ be a KT manifold with metric $g$ and complex structure $J$.
We in addition assume that the gauge group $G$ acts on $M$ with
isometries which furthermore preserve the complex structure $J$ and
the Wess-Zumino term $H$. These conditions are the same as those in the case
of (2,0)-supersymmetric theory. The (2,1) sigma model superfield $\phi$
is a map from the (2,1) superspace $\Xi^{2,1}$ into the sigma model
manifold $M$.  In addition it is required that
\begin{align}
\nabla_{1+}\phi^i= J^i{}_j \nabla_{0+}\phi^j\ ,
\label{s21con}
\end{align}
where $\nabla_{p+}\phi^i=D_{p+}\phi^i+ A^a_{p+} \xi^i_a$ and $\xi_a$
are the vector fields on $M$ generated by the group action. As we have
seen, the (2,1) gauge multiplet satisfies the condition \eqref{g21con}
similar to \eqref{s21con}.  The superfield $\phi$ is also covariantly
chiral, as can be seen by choosing complex coordinates on the sigma
model manifold $M$. These results will be used later for the
construction of actions of (2,1)-supersymmetric gauge theories coupled
to sigma models.

The components of the sigma model multiplet $\phi$ are as follows:
\begin{align}
  \phi^i &= \phi^i|
& \ell^i &= \nabla_{0+}\nabla_-\phi^i|\nonumber \\
  \lambda^i_{+}=&\nabla_{0+}\phi^i|
& \lambda^i_{-}=&\nabla_{-}\phi^i| ~,
\end{align}
where $\phi$ is a scalar, $\lambda_+$ and $\lambda_-$ are real fermions, and $\ell$ is an auxiliary
field.

\subsection{Action}

An action of a (2,1)-supersymmetric gauge theory coupled to sigma model matter can be written
as
\begin{align}
  S &= S_g + S_\sigma+S_p  \ ,
\end{align}
where $S_g$ is the action of the gauge multiplet, $S_\sigma$ is the
action of the sigma-model multiplet and $S_p$ contains the potential term.
 We shall describe each term separately.

\subsection{The gauge multiplet action}

An action for the (2,1)-supersymmetric gauge multiplet
 is
\begin{align}
\label{eq:cpt21action}
  \begin{split}
    S_g=\intd{^2x\rd\theta_{0}^+\rd\theta^-}  \bigl(&
       - u^0_{ab}\delta^{pq} \nabla_{0+}W_p^a \nabla_- W_q^b
        \\
&     + u^1_{ab} \delta^{pq} \nabla_{1+}W_p^a \nabla_- W_q^b+ z^p_a W^a_p
  \bigr)
  \end{split}
\end{align}
where $u^0, u^1$ are the gauge coupling constants  and $z^p$ are theta term type of couplings. All the
couplings are allowed to depend on the superfield $\phi$. We shall assume that
both $u^0, u^1$ are symmetric in the gauge indices but this
restriction can be lifted.

Observe that the action \eqref{eq:cpt21action}
 is not an integral over the full $\Xi^{2,1}$
superspace.  Therefore it is not manifestly (2,1)-supersymmetric.  The
requirement of invariance of the action under (2,1) supersymmetry imposes the
conditions
\begin{align}\begin{split}
J^j{}_i\partial_j u^0_{ab}&= -\partial_i u^1_{ab}
\\
J^j{}_i\partial_j z^1_{a}&= -\partial_i z^0_{a}
 \ .
\end{split}\end{align}
Therefore the couplings $u^0+i u^1$ and $z^1+i z^0$ are holomorphic.

In addition gauge invariance of the action \eqref{eq:cpt21action} implies that
\begin{align}\begin{split}
{\cal L}_a u^p_{bc}&= -f^d{}_{ab} u^p_{dc}- f^d{}_{ac} u^p_{bd}
\\
{\cal L}_a z^p_{b}&= -f^d{}_{ab} z^p_{d}
\ .
\end{split}\end{align}

\subsection{The sigma model multiplet action and potential}

The action of the (2,1)-supersymmetric gauged sigma model with Wess-Zumino
term has been given in
\cite{hgps}.  Here we shall summarise some of results
relevant to this paper. The action of this multiplet is
\begin{align}
\begin{split}
  S_\sigma =& \intd{^2x\rd\theta_{0}^+\rd\theta^-}
    \big(g_{ij} \nabla_{0+}\phi^i \nabla_-\phi^j+\nu_a W^a_1 \bigr) \\
  &+ \intd{^2x\rd t\rd\theta^+\rd\theta^-}\bigl(
  H_{ijk} \partial_t\phi^i \nabla_{0+}\phi^j \nabla_-\phi^k -
  w_{ia} \partial_t\phi^i W_-^a
  \bigr)
\end{split}
\end{align}
This action can be written without the $t$ integration as
\begin{align}
\begin{split}
  S_\sigma = \intd{^2x\rd\theta^+\rd\theta^-}\bigl(&
    g_{ij} \nabla_+\phi^i \nabla_-\phi^j
    + b_{ij} D_+\phi^i D_-\phi^j \\
&   - A_+^a w_{ia} D_-\phi^i
    - A_-^a w_{ia} D_+\phi^j
    + A_-^a A_+^b \xi_{[b}^i w_{a]i}
  \bigr)
\end{split}
\end{align}

Gauge invariance of the above action requires that $w$ should satisfy the
conditions described in section \ref{sub:gaugecon}. 
As in the case of (2,0)-supersymmetric
gauged sigma model, it is also required that $\nu$ is
globally defined and $\mathcal{L}_a \nu_b =- f_{ab}^{\phantom{ab}c} \nu_c $.
 In fact $\nu$ is a moment map
associated with the action of the gauge group on the KT manifold $M$.

The part of the action involving the potential is
\begin{align}
S_p=\intd{^2x \rd\theta_{0}^+ \rd\theta^{-}} h\ ,
\end{align}
where $h=h(\phi)$
Invariance under (2,1) supersymmetry requires that
\begin{align}
\partial_i h= J^k{}_i \partial_k h^1
\end{align}
where $h^1 =h^1(\phi)$. This implies that $h$ is the real part of a holomorphic function on $M$.

The scalar potential of (2,1)-supersymmetric gauge theories  coupled to sigma models described above is
\begin{align}
V={1\over4}  u_0^{ab} (\nu_a+ z_a^1) (\nu_b+z_b^1)+ {1\over4} g^{ij} (\partial_i h+\partial_i z^p_a W^a_p)
 (\partial_j h++\partial_j z^p_a W^a_p)\ .
\label{morepot}
\end{align}

\subsection{A generalisation}

As we have shown both the (2,1) gauge multiplet 
and the (2,1) sigma model multiplet
are constructed from covariantly chiral scalar 
superfields, ie both satisfy the  conditions
\eqref{s21con} and \eqref{g21con}. Because of this, 
these two superfields can be
combined to a single superfield $Z=( W_0, W_1, \phi)$ which
is a map from the $(2,1)$ superspace $\Xi^{2,1}$
into $({\cal L}G\otimes \bR^2)\times M$. 
In addition we can take $Z$ to satisfy
a chirality condition which is the 
combination of \eqref{g21con} and \eqref{s21con}.
Next we can take the gauge group $G$ 
to act on $({\cal L}G\otimes \bR^2)\times M$
with the adjoint action in the first factor and a group action on $M$. The
vector fields associated by such a group action are
\begin{align}
\xi_a= \sum_p f^c{}_{ab} W_p^b 
{\partial\over\partial W^c_p} +\xi^i \partial_i\ .
\end{align}

Treating the (2,1)-supersymmetric gauge theory coupled to sigma model matter as
a sigma model with superfield $Z$, which
 satisfies \eqref{g21con} and \eqref{s21con},
we can write the action
\begin{align}
\begin{split}
  S_\sigma =& \intd{^2x\rd\theta_{0}^+\rd\theta^-}
   \big( g_{AB} \nabla_{0+}Z^A \nabla_-Z^B+ \nu_a W^a_1+ h \big)\\
  &+ \intd{^2x\rd t\rd\theta^+\rd\theta^-}\bigl(
  H_{ABC} \partial_tZ^A \nabla_{0+}Z^B \nabla_-Z^C -
  w_{Ba} \partial_tZ^B W_-^a
  \bigr)
\end{split}
\label{g21act}
\end{align}
where now all the couplings are defined 
using the geometry of $({\cal L}G\otimes \bR^2)\times M$.
Of course (2,1)-supersymmetric 
requires that $({\cal L}G\otimes \bR^2)\times M$ is a
KT manifold with respect to the 
complex structure $(J, {\rm id}\otimes \epsilon)$.
In particular the metric and the rest 
of the couplings depend on the coordinates of
$({\cal L}G\otimes \bR^2)\times M$. The conditions 
for gauge invariance are easily
determined from those of the (2,1)-supersymmetric gauged sigma model.
We remark that the couplings of \eqref{g21act} 
can be arranged such that the $SO(2)$ R-symmetry of the
$(2,1)$-supersymmetry algebra is broken. In particular 
the $SO(2)$ rotation that rotate
the $W_p$ scalar components is not a symmetry of the action. 
However if one insists in preserving
the R-symmetry, then the KT manifold 
$({\cal L}G\otimes \bR^2)\times M$ should admit
a $SO(2)$ action preserving all the geometric data.

\section{$(4,1)$ supersymmetry}

\subsection{The gauge multiplet}

The $(4,1)$ superspace $\Xi^{4,1}$ has coordinates
$(x^\plpl,x^\mimi,\theta_{p}^+,\theta^-)$, 
where $(x^\plpl,x^\mimi)$ are the even and $\{\theta^-,\theta_{p}^+,
p=0,\ldots,3\}$ are the odd coordinates.
The (4,1)-supersymmetric Yang-Mills multiplet is described by a
connection $A$ in $\Xi^{4,1}$ superspace with components
$(A_\plpl,A_\mimi,A_{p+},A_-)$ with $p=0,\ldots 3$. In addition it is
required that these satisfy the supersymmetry constraints \cite{hgps}
\begin{gather*}
  \begin{align}
  [ \nabla_{p+}, \nabla_- ] &= W_p
& [ \nabla_\plpl, \nabla_\mimi ] &=  F_{\plpl\mimi} \nonumber\\
  [ \nabla_{p+}, \nabla_{q+} ] &= 2i \delta_{pq} \nabla_\plpl
& [ \nabla_-, \nabla_- ] &=  2i\nabla_\mimi
\end{align}  \label{con41v}  
\\
  \nabla_{p+} W_q = \epsilon_{pq}^{\phantom{pq}p'q'} \nabla_{p'+} W_{q'}\ .
\end{gather*}
(We have suppressed all the gauge indices.)
The Jacobi identities imply that
  \begin{gather*}
  \begin{align}
    [\nabla_{p+}, \nabla_\mimi ] &= i\nabla_-W_p \nonumber\\
    [\nabla_-, \nabla_\plpl ] &= i\nabla_{0+} W_0 \nonumber\\
    F_{\plpl\mimi}^a &= \nabla_{0+} \nabla_- W_0^a   
  \end{align}  \label{jac41con}
\\
\nabla_{p+} W_{q}+ \nabla_{q+} W_p=0 \quad (p\neq q) \nonumber
\\
\nabla_{0+} W_{0}=\nabla_{1+} W_{1}= \nabla_{2+} W_{2}=\nabla_{3+}
W_{3}
  \end{gather*}

The (4,1) gauge multiplet is determined 
by four scalar superfields. Some of the conditions
on these superfields given in \eqref{jac41con}, 
like in the (2,1) model previously, can be
expressed as conditions of a (4,1) sigma model multiplet.  For this,
 view the four-scalar
superfields $\{W_p: p=0,1,2,3\}$ as maps from
 the superspace $\Xi^{4,1}$ into ${\cal L}G\otimes \bR^4$, where ${\cal L}G$
is the Lie algebra of the gauge 
group $G$. Then introduce three constant complex structures $\{I_r \}$
in $\bR^4$ such that $(I_r)^0{}_s=\delta_{rs}$ 
and $(I_r)^s{}_t=- \epsilon_{rst}$ where
$r,s,t=1,2,3$. The conditions on $W_p$ in \eqref{con41v} 
and \eqref{jac41con} can be expressed
as
\begin{align}
\nabla_{r+} W^a_p= I_r{}^q{}_p \nabla_{0+}W^a_q\ .
\label{g41con}
\end{align}

The components of the gauge multiplet are
\begin{gather}
\begin{align}
  W_p &= W_p| 
& F_{\plpl\mimi} &= \nabla_{0+}\nabla_-W_0| \nonumber\\
  \chi_{p-} &= \nabla_- W_p| 
& \chi_{p+} &= \nabla_{0+} W_p |
 \end{align}
 \\
 f_r = \nabla_{0+}\nabla_- W_r| \quad r=1,2,3\ , \nonumber
\end{gather}
where $W_p$ are scalars, $\chi_{p+}$,$\chi_{p-}$ are the
gaugini which are real chiral fermions in two dimensions,
$F_{\plpl\mimi}$ is the field strength and $\{f_{r}: r=1,2,3\}$ are auxiliary
fields. The $SO(4)$ R-symmetry of the (4,1)-supersymmetric gauge theory rotates
both the scalars and the fermions of the gauge multiplet.

\subsection{The sigma model multiplet}

Let $M$ be a HKT manifold with metric 
$g$ and hypercomplex structure $\{J_r ; r=1,2,3\}$.
We in addition assume that the 
gauge group $G$ acts on $M$ with isometries which in
addition preserve the hypercomplex 
structure $J_r$ and the Wess-Zumino term $H$. These
conditions are the same as in the case of (4,0)-supersymmetric
 model. The (4,1) sigma model
superfield $\phi$ is a map  from the
 (4,1) superspace $\Xi^{4,1}$ into the sigma model manifold $M$.
In addition it is required that
\begin{align}
\nabla_{r+}\phi^i= J_r{}^i{}_j \nabla_{0+}\phi^j\ ,
\label{s41con}
\end{align}
where $\nabla_{p+}\phi^i=D_{p+}\phi^i+ A^a_{p+} \xi^i_a$ 
and $\xi$ are the vector
fields on $M$ generated by the group action. As we have seen the
(4,1) gauge multiplet satisfies 
the condition \eqref{g41con} similar to \eqref{s41con}.
 These results
will be used later for the construction of actions of (4,1)-supersymmetric
gauge theories coupled to sigma models.

The components of the sigma model (4,1) multiplet $\phi$ are as follows,
\begin{align}
  \phi^i &= \phi^i|
& \ell^i &= \nabla_{0+}\nabla_-\phi^i| \nonumber\\
  \lambda^i_{+}&=\nabla_{0+}\phi^i|
& \lambda^i_{-}&=\nabla_{-}\phi^i| ~,
\end{align}
where $\phi$ is a scalar, $\lambda_+$ and $\lambda_-$ 
are real fermions, and $\ell$ is an auxiliary
field.

\subsection{Action}

The action of a (4,1)-supersymmetric gauged theory coupled to sigma model
matter can be written
as
\begin{align}
  S &= S_g + S_\sigma+ S_p  \ ,
\end{align}
where $S_g$ is the action of the gauge multiplet, $S_\sigma$ is the
action of the sigma-model multiplet and $S_p$ is the potential. We
shall describe each term separately.

\subsection{The gauge multiplet action}

An action for the (4,1)-supersymmetric gauge multiplet
 is
\begin{align}
\label{eq:cpt41action}
  \begin{split}
    S_g=\intd{^2x\rd\theta_{0}^+} \bigl(&
       - u^0_{ab}\delta^{pq} \nabla_{0+}W_p^a \nabla_- W_q^b
        \\
&     + u^r_{ab} \delta^{pq} \nabla_{r+}W_p^a \nabla_- W_q^b+ z_a^p W^a_p
  \bigr)
  \end{split}
\end{align}
where $\{u^p\}=\{u^0, u^r\}$ and $z^p$ are the gauge coupling
constants and theta type of terms, respectively, which in general
depend on the superfield $\phi$. We shall assume that both $u^p$ are
symmetric in the gauge indices but this restriction can be lifted.

Observe that this action is not an integral over the full $\Xi^{4,1}$
superspace.  Therefore it is not manifestly (4,1)-supersymmetric.  The
requirement of invariance under (4,1) supersymmetry imposes the
condition
that $u^p$ and $z^p$ are constant. This is similar 
to the condition that arises in
(4,0) supersymmetric gauge theories. In addition, gauge invariance 
of the action
\eqref{eq:cpt41action}  requires that
\begin{align}\begin{split}
 f^d{}_{ab} u^p_{dc}+ f^d{}_{ac} u^p_{bd}&=0
 \\
  f^d{}_{ab} z^p_d&=0
 \ .
\end{split}\end{align}
Thus $u^p$ must be invariant quadratic forms on the Lie algebra of
the group $G$ and $z^p$ must be  invariant elements of the Lie algebra.
Of course $z^p=0$, if $G$ is semi-simple.

\subsection{The sigma model multiplet action and the potential}

An action for the (4,1) sigma model multiplet coupled to gauge fields
has been given in
\cite{hgps}.  Here we shall summarise some of results
relevant to this paper.  The action of the (4,1)-supersymmetric
gauged sigma model is
\begin{align}
\begin{split}
  S_\sigma =& \intd{^2x\rd\theta^+\rd\theta^-}
   \big( g_{ij} \nabla_+\phi^i \nabla_-\phi^j+\sum_r \nu^r_a W^a_r\big) \\
  &+ \intd{^2x\rd t\rd\theta^+\rd\theta^-}\bigl(
  H_{ijk} \partial_t\phi^i \nabla_+\phi^j \nabla_-\phi^k -
  w_{ia} \partial_t\phi^i W_-^a
  \bigr)\ .
\end{split}
\end{align}

The gauge transformations of $\phi$ are $\delta\phi^i =
\lambda^a\xi_a^{\phantom{a}i}(\phi)$.  Gauge invariance of the above
action requires that $w$ should satisfy the conditions described in
section \ref{sub:gaugecon}. As in the case of (4,0)-supersymmetric
gauged sigma model, $\nu^r$ should satisfy $\mathcal{L}_a \nu^r_b =-
f_{ab}^{\phantom{ab}c} \nu^r_c $. In fact $\nu^r$ is a moment map
associated with the action of the gauge group $G$ on the HKT manifold
$M$.

The part of the action involving the potential is
\begin{align}
S_p=\intd{^2x \rd\theta_{0}^+ \rd\theta^{-}} h\ ,
\end{align}
where $h=h(\phi)$.
Invariance under (4,1) supersymmetry requires that
\begin{align}
\partial_i h= J_r{}^k{}_i \partial_k h^r
\end{align}
where $h^r =h^r(\phi)$. This implies that $h$ is the real
part of three holomorphic functions on $M$, ie $h$
is tri-holomorphic.

The scalar potential of (4,1)-supersymmetric
 gauge theories coupled to sigma models is
\begin{align}
V={1\over4}  u_0^{ab}\sum_{r=1}^3 \nu^r_a \nu^r_b
+ {1\over4} g^{ij} \partial_i h \partial_j h\ ,
\label{more41pot}
\end{align}
where we have shifted the moment maps $\nu_r$ by a constant $z_r$.

\subsection{A generalisation}

As we have shown both the (4,1) gauge multiplet and the (4,1) sigma
model multiplet are constructed from scalar superfields which satisfy
the similar constraints \eqref{g41con} and \eqref{s41con}, respectively.
 Because of
this, these two superfields can be combined to a single superfield
$Z=( W, \phi)$ which is a map from the $(4,1)$ superspace $\Xi^{4,1}$
into $({\cal L}G\otimes \bR^4)\times M$. In addition we can take $Z$
to satisfy a condition which is the combination of \eqref{s41con} and
\eqref{g41con}.  Next we can take the gauge group $G$ to act on
$({\cal L}G\otimes \bR^4)\times M$ with the adjoint action in the
first factor and a group action on $M$. The vector fields associated
by such a group action are
\begin{align}
\xi_a= \sum_p f^c{}_{ab} W_p^b {\partial\over\partial W^c_p}
 +\xi^i \partial_i\ .
\end{align}

Treating the (4,1)-supersymmetric gauge theory coupled to sigma model
matter as a sigma model with superfield $Z$, which satisfies
\eqref{g41con} and \eqref{s41con}, we can write the action
\begin{align}
\begin{split}
  S_\sigma =& \intd{^2x\rd\theta_{0}^+\rd\theta^-}
   \big( g_{AB} \nabla_{0+}Z^A \nabla_-Z^B+\sum_r \nu^r_a W^a_r+ h \big)\\
  &+ \intd{^2x\rd t\rd\theta^+\rd\theta^-}\bigl(
  H_{ABC} \partial_tZ^A \nabla_{0+}Z^B \nabla_-Z^C -
  w_{Ba} \partial_tZ^B W_-^a
  \bigr)
\end{split}
\label{g41act}
\end{align}
where now all the couplings are defined using the geometry of $({\cal
  L}G\otimes \bR^4)\times M$.  Of course (4,1)-supersymmetry requires
that $({\cal L}G\otimes \bR^4)\times M$ is a HKT manifold with respect
to the hypercomplex structure $(J_r, I_r)$.  In particular the metric
and the rest of the couplings depend on the coordinates of $({\cal
  L}G\otimes \bR^4)\times M$. The conditions for gauge invariance are
easily determined from those of the (4,1)-supersymmetric gauged sigma
model.  We remark that the couplings of \eqref{g41act} can be arranged
such that the $SO(4)$ R-symmetry of the $(4,1)$-supersymmetry algebra
is broken. In particular the $SO(4)$ rotation that rotates the $W_p$
scalar components is not a symmetry of the action. However if one
insists in preserving the R-symmetry, then the HKT manifold $({\cal
  L}G\otimes \bR^4)\times M$ should admit a $SO(4)$ action preserving
all the geometric data.

\section{A bound for vortices in the (2,0) model}

Vortices are the instantons of two-dimensional gauge theories coupled
to sigma models.  Bogomol'nyi type of bounds for both abelian
\cite{bog} and non-abelian vortices \cite{matha, mathb} have been
investigated in the context of linear sigma models.  Here we shall
establish bounds for vortices for non-linear sigma models.  For this
we shall consider the Euclidean action of the (2,0)-supersymmetric
gauge sigma model without Wess-Zumino term. The sigma model target
space $M$ is K\"ahler with metric $g$, complex structure $J$ and
associated K\"ahler form $\Omega_J$ ($(\Omega_J)_{ij}=g_{ik} J^k{}_j$).
 After a Wick rotation the
two-dimensional spacetime is $\bR^2$ with the standard Euclidean
metric.  The relevant part of the bosonic Euclidean action of a
(2,0)-supersymmetric gauge theory coupled to a sigma model is
\begin{align}
  S_E = \ints{\bR^2}{\rd^2x} 
  \bigl(\frac{1}{2} g_{ij}\delta^{\mu\nu} \nabla_\mu \phi^i  \nabla_\nu \phi^j
+ \frac{1}{2} u_{ab} F^a_{\mu\nu} F^b_{\lambda\rho}
 \delta^{\mu\lambda} \delta^{\nu\rho}
+\frac{1}{4} u^{ab} \nu_a \nu_b\bigr)\ .
\label{eu20act}
\end{align}
Next we introduce  $I$ a constant complex structure  on $\bR^2$ such that
$\bR^2$ is a K\"ahler manifold. The associated K\"ahler form $\Omega_I$ is the volume form of $\bR^2$.
In such a case the Euclidean action \eqref{eu20act}
can be rewritten as
\begin{align}
\begin{split}
  S_E=&\intd{^2x} \bigl[
    \frac{1}{4} u_{ab} \bigl((\Omega_I \cdot F^a\mp {1\over2}\nu^a)
    (\Omega_I \cdot F^b\mp {1\over2}\nu^b)
\\
&+ {1\over4} g_{ij} \delta^{\mu\nu}(I^\rho{}_\mu\nabla_\rho\phi^i
\mp \nabla_\mu \phi^k J^i{}_k)
(I^\sigma{}_\nu\nabla_\sigma\phi^j\mp \nabla_\nu \phi^\ell J^j{}_\ell)] \\
&\pm \int_{\bR^2} ((\Omega_J)_{ij} \nabla\phi^i\wedge \nabla\phi^j +\nu_a F^a)
\end{split}
\label{bogsquare}
\end{align}
where $\Omega_I\cdot F=(\Omega_I)^{\mu\nu} F_{\mu\nu}$, 
$\nu^a=u^{ab} \nu_b$, $u_{ab}=u^0_{ab}$ and
$u^{ac}u_{cb}=\delta^a{}_b$ ($u_{ab}=u_{(ab)}$). We
 remark that the above expression
for the Euclidean action has been constructed from
 \eqref{eu20act} by completing squares and collecting
all the remaining terms which organise
 themselves in the last term of \eqref{bogsquare}.

The last term in \eqref{bogsquare},
\begin{align}
 {\cal Q}=\int_{\bR^2}  \omega_J\ ,
 \end{align}
 is a topological charge, where the form
 \begin{align}
 \omega_J=(\Omega_J)_{ij} \nabla\phi^i\wedge \nabla\phi^j +\nu_a F^a
\end{align}
is the {\it equivariant extension of the K\"ahler form} $\Omega_J$ of
the sigma model target space $M$.  The form $\omega_J$ is closed.
Viewing $\omega_J$ as form on $\bR^2$, it is apparent.  In fact
$\omega_J$ is closed as a two-form on any manifold $N$ for any map
$\phi$ from $N$ into the sigma model manifold $M$ and for any choice
of connection $A$.  This can be easily seen and we shall not
demonstrate it here.

The Euclidean action of the (2,0)-supersymmetric two-dimensional
gauge theory coupled to a sigma
model is bounded by the absolute value of the topological charge
${\cal Q}$, $S_E\geq |{\cal Q}|$. This is because it is always
possible to choose the signs in the Bogomol'nyi bound above such that
the topological term is positive. If the topological charge is
positive, then the bound is attained whenever
\begin{align}\begin{split}
\Omega_I\cdot F^a-\nu^a&=0
\\
J^i{}_j\nabla_\mu\phi^j-\nabla_\nu \phi^i I^\nu{}_\mu&=0\ .
\end{split}\end{align}
In two-dimensions, the curvature $F$ is a (1,1)-form.  Choosing
complex coordinates $(z, \bar z)$ on $\bR^2$ with respect to the
complex structure $I$, it is always possible to arrange using a
(complex) gauge transformation that $A_{\bar z}=0$. Choosing complex
coordinates in the sigma model target space $M$ as well, it is easy to
see that the second BPS condition implies that the map $\phi$ is
holomorphic from the spacetime $\bR^2$ into the sigma model manifold
$M$.

A special case of this bound arises for gauge theories couple to linear
sigma models for which the sigma model manifold $M=\bR^{2n}$ with the
Euclidean metric and equipped with a constant compatible complex
structure $J$. This case includes the Nielsen-Olesen vortices
\cite{hnpo}. (For these, existence of a solution was shown in
\cite{taubes} and the moduli were studied in \cite{weiberg},
\cite{matha} and more recently in \cite{manton}, see also
\cite{nekrasov}).  The case with a single complex scalar has been
analysed in \cite{bog}.  Choosing complex coordinates $\{q^\alpha;
\alpha=1,\dots,n\}$ in $\bR^{2n}$, we write
\begin{align}\begin{split}
\rd s^2&=\sum_{\alpha} \rd q^\alpha \rd q^{\bar\alpha}
\\
\Omega_J&=-i \sum_{\alpha} \rd q^{\alpha}\wedge  \rd q^{\bar\alpha}\ .
\end{split}\end{align}
Next consider the abelian group $U(1)$-action $q^\alpha\rightarrow
e^{i Q_\alpha t} q^\alpha$ which generates the holomorphic Killing
vector fields
\begin{align}
\xi=i \sum_{\alpha} Q_\alpha
 (q^\alpha {\partial\over \partial q^\alpha}-q^{\bar \alpha}
{\partial\over \partial q^{\bar\alpha}})
 \ .
\end{align}
 The moment map is
\begin{align}
\nu=- \sum_{\alpha}( Q_\alpha q^\alpha q^{\bar\alpha})-\Lambda\ ,
\end{align}
where $\Lambda$ is a (cosmological) constant. This is an example of a
$(2,0)$-supersymmetric gauged linear sigma model with gauge group
$U(1)$ of the type considered in \cite{witten}.  The topological
charge is
\begin{align}
{\cal Q}=\ints{\bR^2}{\rd^2z}
 \big(\sum{\alpha}(\nabla_z q^\alpha \nabla_{\bar z} q^{\bar \alpha}-
\nabla_{\bar z} q^\alpha \nabla_{ z} q^{\bar \alpha})+ \nu F_{z\bar z}\big)
\label{topcharge}
\end{align}
where $\nabla_z q^\alpha=\partial_z q^\alpha+i A_z Q_\alpha q^\alpha$,
$\nabla_z q^{\bar \alpha}=\partial_z q^{\bar\alpha}
-i A_z Q_\alpha q^{\bar\alpha}$,
$\nabla_{\bar z} q^{\bar \alpha}=(\nabla_z q^\alpha)^*$ and
$\nabla_{\bar z} q^{\alpha}=(\nabla_z q^{\bar \alpha})^*$, and
$F_{z\bar z}=\partial_z A_{\bar z}- \partial_{\bar z} A_z$. 
To compare the bound above \eqref{bogsquare} with that of vortices in
\cite{witten}, we observe that {\it after some integration by parts}
we have
\begin{align}
{\cal Q}=\int_{\bR^2} d^2z \big(\sum_{\alpha}
 Q_\alpha (\partial_z q^\alpha \partial_{\bar z} q^{\bar \alpha}-
\partial_{\bar z} q^\alpha \partial_{ z}
 q^{\bar \alpha})-\Lambda F_{z\bar z}\big)+ {\rm surfaces}
\end{align}
The first term in the above expression is the topological charge
expected for the vortices (instantons) of ungauged two-dimensional
sigma models. The same topological charge also appears in the kink
solitons of three-dimensional non-linear sigma models. The last part
in the above expression involving the cosmological constant and the
Maxwell field is the usual degree of an abelian vortex.  The relation
between the topological charge ${\cal Q}$ in \eqref{topcharge} and the
degree of an abelian vortex involves integration by parts. Under
certain boundary conditions the two topological charges are the same.
However as we have shown, the bound that involves the equivariant
extension of the K\"ahler form generalizes in the context of gauge
theories coupled to non-linear sigma models.

\section{A bound for vortices in the (4,0) model}

A bound similar to the one we have described in the previous section
for the Euclidean action of (2,0)-supersymmetric gauge theory coupled
to sigma model matter 
can also be found for the Euclidean action of (4,0)-supersymmetric
gauge theory.  The Euclidean action of a (4,0)-supersymmetric
gauge theory coupled to sigma model matter with vanishing Wess-Zumino term is
\begin{align}
S_E= \ints{\bR^2}{\rd^2x} \big({1\over2}
 g_{ij}\delta^{\mu\nu} \nabla_\mu \phi^i  \nabla_\nu \phi^j
+ {1\over2} u_{ab} F^a_{\mu\nu} F^b_
{\lambda\rho} \delta^{\mu\lambda} \delta^{\nu\rho}
+{1\over4} \sum_{r=1}^3 u^{ab}\sum_{r} \nu^r_a \nu^r_b\big)
\end{align}
The sigma model target space is hyper-K\"ahler with metric $g$,
hypercomplex structure $\{J_r: r=1,2,3\}$ and associated K\"ahler
forms $\Omega_{J_r}$.  After the Wick rotation, the two-dimensional
spacetime is $\bR^2$ with the standard Euclidean metric.  Let $I$ be a
compatible constant complex structure such that $\bR^2$ is a K\"ahler
manifold with associated K\"ahler form $\Omega_I$. In such a case the
Euclidean action can be written as
\begin{align}
\begin{split}
S_E=&\intd{^2x} [{1\over4}\sum_{r=1}^3 u_{ab} 
(a_r \Omega_{I}\cdot F^a\mp \nu_r^a )
( a_r \Omega_{I}\cdot F^b\mp \nu_r^b)
\\
&+ {1\over4}\sum_{r=1}^3 \delta^{\mu\nu} g_{ij}
 ( a_r I^\rho{}_\mu\nabla_\rho\phi^i\mp \nabla_\mu \phi^k J_r^i{}_k)
( a_r I^\sigma{}_\nu\nabla_\sigma\phi^i\mp \nabla_\nu \phi^\ell J_r^j{}_\ell)]
\\ & \pm
\int_{\bR^2} \sum_{r=1}^3 a_r \omega_{J_r}\ ,
\end{split}
\label{bou40}
\end{align}
where $\{a_r :r=1,2,3\}$ is a 
constant vector with length one, $\sum_{r=1}^3 (a_r)^2=1$,
$u_{ab}=u^0_{ab}=u^0_{ba}$, $\nu^a_r=
u^{ac} \nu_{rc}$, $u^{ac} u_{cb}=\delta^a{}_b$ and
\begin{align}
\omega_{J_r}=(\Omega_{J_r})_{ij} \nabla\phi^i\wedge \nabla\phi^j +\nu_a F^a
\end{align}
is the equivariant extension of the K\"ahler form $\Omega_{J_r}$.

The strictest  bound is attained whenever
 the unit vector $\{a_r: r=1,2,3\}$ is parallel
to the vector of the topological charges ${\cal Q}_r: r=1,2,3\}$, where
\begin{align}
{\cal Q}_r=\int_{\bR^2} \omega_{J_r}
\end{align}
and the sign is chosen such that the topological term in the bound is
positive.  If the inner product of $\{a_r: r=1,2,3\}$ and $\{{\cal
  Q}_r: r=1,2,3\}$ in \eqref{bou40} is positive, we have that
\begin{align}
S_E\geq \sqrt {{\cal Q}_1^2+{\cal Q}_2^2+{\cal Q}_3^2}\ .
\end{align}
This bound is attained whenever
\begin{align}
\begin{split}
 a_r \Omega_{I}\cdot F^a- \nu_r^a&=0
\\
{J_r}^i{}_j\nabla_\mu\phi^j-a_r \nabla_\nu \phi^i I^\nu{}_\mu&=0\ .
\end{split}
\label{bog40eqn}
\end{align}

Using a rotation in the space of three complex structures, we can
arrange such that $a_1=1$ and $a_2=a_3=0$.  In such case, the last
equation in \eqref{bog40eqn} implies that
\begin{align}
\nabla_\mu \phi^i=0\ .
\end{align}
This in turn gives
\begin{align}
F_{\mu\nu}^a \xi_a^i=0
\end{align}
Therefore either $\phi$ takes values in the fixed point set $M_f$ of
the group action of $G$ in $M$ or the curvature $F$ of the connection
$A$ vanishes. In the latter case, the first equation in
\eqref{bog40eqn} implies that $\nu_r=0$ and these are the vacua of the
theory. If these are no non-trivial flat connections and $M_f\cap
\nu_1^{-1}(0)\cap \nu_2^{-1}(0)\cap \nu_3^{-1}(0)$ is empty, then the
space of solutions if the hyper-K\"ahler reduction $M//G$ of $M$ and
it is a hyper-K\"ahler manifold. On the other hand if $\phi$ take
values in $M_f$, then the second equation in \eqref{bog40eqn} implies
that $\phi$ are constant. Substituting this in the first equation in
\eqref{bog40eqn} implies that $\phi$ are in $M_f\cap \nu_2^{-1}(0)\cap
\nu_3^{-1}(0)$. In addition we have that
\begin{align}
\Omega_{I}\cdot F^a- \nu_1^a=0\ .
\end{align}
This is the Hermitian-Einstein equation in two dimensions.

\section{K\"ahler manifolds and  non-abelian vortices}

The bounds that we have described in the previous sections can be
easily generalized as follows. Consider two K\"ahler manifolds
$(N,h,I)$ and $(M,g,J)$ of dimensions $2k$ and $2n$, and K\"ahler
forms $\Omega_I$ and $\Omega_J$, respectively. Next allow $M$ to admit
a holomorphic $G$-action with associated killing vector fields $\xi$
and moment map $\nu$. In our conventions $i_{\xi}\Omega_J=-d\nu$.  Next
consider the functional
\begin{align}
S_E=\int_N d{\rm vol}(N) \big({1\over2}
|\nabla\phi|^2+ {1\over2} |F|^2+{1\over4} |\nu|^2\big)\ ,
\end{align}
where $|\nabla|^2= g_{ij} h^{\mu\nu} \nabla_\mu\phi^i \nabla_\nu \phi^j$,
$\nabla_\mu\phi^i=\partial_\mu \phi^i+ A^a \xi_a^i$,
$|F|^2=u_{ab} F^a_{\mu\nu} F^b_{\rho\sigma} h^{\mu\rho} h^{\nu\sigma}$, $|\nu|^2= u^{ab} \nu_a \nu_b$
and $u$ is a fibre inner product on the gauge bundle which we can set $u_{ab}=\delta_{ab}$.

The functional $S_E$ can be rewritten as follows:
\begin{align}
\begin{split}
S_E=&\int_N d{\rm vol}(N) \big[ {1\over 4} |\Omega_I\cdot F\mp \nu|^2+ |F^{2,0}|^2
+ {1\over 4} |I\nabla\phi\mp J\nabla\phi|^2\big]
\\
&\pm {1 \over (k-1)!}\int_N \omega_J\wedge \Omega_I^{k-1}
-{1 \over (k-2)!}\int_N  u_{ab} F^a\wedge F^b\wedge \Omega_I^{k-2}
\end{split}
\label{bogkk}
\end{align}
where we have chosen the normalisation $d{\rm vol}(N)={1\over k!} \Omega_I^k$,
$\Omega_I\cdot F=\Omega_I^{\mu\nu} F_{\mu\nu}$, $F^{2,0}$ is the (2,0) part of the curvature $F$ and
\begin{align}
\omega_J=(\Omega_J)_{ij} \nabla\phi^i\wedge \nabla\phi^j+ \nu_a F^a
\end{align}
is the equivariant extension of the K\"ahler form $\Omega_J$. (The
inner products are taken with respect to the Riemannian metrics $h$
and $g$.) The rest of the notation is self-explanatory.  We remark
that if $\Omega_J$ represents the first Chern class of a line bundle,
i.e.\  the K\"ahler manifold is Hodge, then $\omega_J$ can be thought of
as the equivariant extension of the first Chern class (see \cite{ab}).

If $u_{ab}$ is a constant invariant quadratic form on the Lie algebra
of the gauge group $G$, it is clear that the functional $S_E$ is
bounded by a topological term ${\cal Q}$ which involves the
equivariant extension of the K\"ahler form and the second Chern
character of the bundle $P\times_G {\cal L}(G)$, where $P$ is a
principal bundle of the gauge group $G$ and $G$ acts on ${\cal L}(G)$
with the adjoint representation.  In particular we can write
\begin{align}
\begin{split}
S_E=&\int_N d{\rm vol}(N) \big[ {1\over 4} |\Omega_I\cdot F\mp \nu|^2+ |F^{2,0}|^2
+ {1\over 4} |I\nabla\phi\mp J\nabla\phi|^2\big]
\\
&\pm {1 \over (k-1)!}\int_N \omega_J\wedge \Omega_I^{k-1}
-{8\pi^2 \over \lambda  (k-2)!}\int_N  ch_2 \wedge \Omega_I^{k-2}\ ,
\end{split}
\label{bogkkk}
\end{align}
where $\lambda$ is an appropriate normalisation factor involving the
ratio between the fibre inner product on $P\times_G {\cal L}(G)$ and
$u$; where $G$ is simple.  It is worth pointing out that the term
involving the second Chern character is not affected by the choice of
sign in writing \eqref{bogkk}.  Therefore there are three cases to
consider the following: (i) there is no choice of sign such that the
topological charge ${\cal Q}$ is positive. In such a case the bound
cannot be attained.  (ii) There is a critical case in which for one
choice of sign the topological charge is negative while for the other
choice is zero.
This case implies that the Euclidean action vanishes and so every term
should vanish. Solutions exist for $F=\nabla\phi=\nu=0$. (iii) For one
of the choice of signs the topological charge is positive.  Suppose
that ${\cal Q}$ is positive in \eqref{bogkk} for the first choice of
sign.  In such case the bound is attained provided that the equations
\begin{align}
\begin{split}
F^{2,0}&=0
\\
\Omega_I\cdot F^a-\nu^a&=0
\\
I^\nu{}_\mu \nabla_\nu\phi^i- J^i{}_j \nabla_\mu \phi^j&=0
\end{split}
\label{bogkkeqn}
\end{align}
hold.  The first equation implies that $F$ is a (1,1)-form. The last
equation in \eqref{bogkkeqn} implies that the maps $\phi$ are
holomorphic. Finally the middle equations are a generalization of non-abelian
vortex equations.  If the term involving the moment map is constant,
then the resulting equation is the Hermitian-Einstein equation.

\section{Hyper-K\"ahler manifolds and  non-abelian vortices}

Let $(N,h,I)$ be a K\"ahler manifold of dimension $2k$ with associated
K\"ahler form $\Omega_I$ and $(M,g,J_r)$ be a hyper-K\"ahler manifold
of dimension $4n$ with associated K\"ahler forms $\Omega_{J_r}$. Next
allow $M$ to admit a tri-holomorphic $G$-action with associated
killing vector fields $\xi$ and  moment maps $\nu_r$.  In our conventions
$i_{\xi}\Omega_{J_r}=-\rd\nu_r$.  Next consider the functional
\begin{align}
S_E=\int_N d{\rm vol}(N) \big({1\over2}
|\nabla\phi|^2+ {1\over2} |F|^2+{1\over4}\sum_{r=1}^3 |\nu_r|^2\big)\ ,
\end{align}
where $|\nabla|^2= g_{ij} h^{\mu\nu} \nabla_\mu\phi^i \nabla_\nu \phi^j$,
$\nabla_\mu\phi^i=\partial_\mu \phi^i+ A^a \xi_a^i$,
$|F|^2=u_{ab} F^a_{\mu\nu} F^b_{\rho\sigma}
 h^{\mu\rho} h^{\nu\sigma}$, $|\nu_r|^2= u^{ab} \nu_{ra} \nu_{rb}$
and $u$ is a fibre inner product on the gauge bundle
 which we can set $u_{ab}=\delta_{ab}$.

The functional $S_E$ can be rewritten as follows:
\begin{align}
\begin{split}
S_E=&\int_N d{\rm vol}(N) \big[ {1\over 4}\sum_{r=1}^3
 |a_r \Omega_I\cdot F\mp \nu_r|^2+ |F^{2,0}|^2
+ {1\over 4}\sum_{r=1}^3 |a_r I\nabla\phi\mp J_r\nabla\phi|^2\big]
\\
&\pm {1 \over (k-1)!}\int_N\sum_{r=1}^3 a_r \omega_{J_r}\wedge \Omega_I^{k-1}
-{1 \over (k-2)!}\int_N  u_{ab} F^a\wedge F^b\wedge \Omega_I^{k-2}
\end{split}
\label{bogkh}
\end{align}
where $d{\rm vol}(N)={1\over k!} \Omega_I^k$, 
$\{a_r: r=1,2,3\}$ is a constant vector of length one,
$\sum_{r=1}^3 (a_r)^2=1$,
$\Omega_I\cdot F=\Omega_I^{\mu\nu} F_{\mu\nu}$,
 $F^{2,0}$ is the (2,0) part of the curvature $F$ and
\begin{align}
\omega_{J_r}=(\Omega_{J_r})_{ij} \nabla\phi^i\wedge \nabla\phi^j+ \nu_{ra} F^a
\end{align}
is the equivariant extension of the K\"ahler form $\Omega_{J_r}$. (The
inner products are taken with respect to the Riemannian metrics $h$
and $g$.) The rest of the notation is self-explanatory.

It is clear that the functional $S_E$ is bounded by a topological
charge ${\cal Q}$ which involves the equivariant extensions of the
K\"ahler forms $\Omega_{J_r}$ and, if $u$ is a constant invariant
quadratic form on ${\cal L}(G)$, the second Chern character of the
gauge bundle $P\times_G {\cal L}(G)$.  It is worth pointing out that
the term involving the second Chern character is not affected by the
choice of sign in writing \eqref{bogkh}. Therefore as in the K\"ahler
case, there are several cases to consider but we shall not repeat the
analysis again. Suppose that both ${\cal Q}$ and that the inner
product of the vector $\{a_r :r=1,2,3\}$ with $\{\tilde{\cal Q}_r :
r=1,2,3\}$ are positive in \eqref{bogkh}, where
\begin{align}
\tilde {\cal Q}_r={1 \over (k-1)!}\int_N  \omega_{J_r}\wedge \Omega_I^{k-1}\ .
\end{align}
 Then the bound is attained provided that the equations
\begin{align}
\begin{split}
F^{2,0}&=0
\\
a_r \Omega_I F^a-\nu_r^a&=0
\\
 a_r I^\nu{}_\mu \nabla_\nu\phi^i- J_r{}^i{}_j \nabla_\mu \phi^j&=0
 \end{split}
\label{bogkheqn}
\end{align}
hold.  The first equation implies that $F$ is a (1,1)-form. It is
always possible with a rotation in the space of complex structures of
the hyper-K\"ahler manifold $M$ to set $a_1=1$ and $a_2=a_3=0$.  Then last
equation in \eqref{bogkheqn} implies that
\begin{align}
\nabla_\mu\phi^i=0\ .
\end{align}
This in turn implies that
\begin{align}
F_{\mu\nu}^a \xi_a^i=0\ .
\end{align}
Therefore either the connection $A$ is flat or the maps $\phi$ take
values in the fixed set $M_f$ of the $G$-group action on $M$. In the
former case, in the absence of non-trivial flat connections, the
moduli space of solutions to these equations is the hyper-K\"ahler
reduction $M//G$ of $G$ and it is a smooth manifold provided that the
level set does not intersect $M_f$.  In the latter case, the maps
$\phi$ are constant and the two remaining equations in
\eqref{bogkheqn} are the Hermitian-Einstein equations for the
connection $A$.

One can also consider the case where $(N,h,I_r)$ is a hyper-K\"ahler
manifold while $(M,g,J)$ is a K\"ahler manifold which admits a
$G$-holomorphic action of isometries.  This case can be treated as
that considered in the previous section involving only K\"ahler
manifolds. A K\"ahler structure on $N$ can chosen with respect to any
complex structure which lies in the two-sphere of complex structures
of $N$.

\section{Concluding Remarks}

We have constructed the actions of two-dimensional (p,0)- and
(p,1)-supersymmetric gauge theories coupled to sigma model matter with
Wess-Zumino term. We have also given the scalar potentials of these
theories. Our method of constructing these theories relies on a
superfield method. Then we have shown that the Euclidean actions these
theories admit vortex type of bounds which generalise to higher
dimensions.

The gauge theories that we have constructed are not the most general
ones. It is known for example that the (1,1)-supersymmetric sigma
model admits a scalar potential which is the length of a killing
vector field \cite{lagf}. Our superfield method cannot describe such
a term. There are also other possibilities, for example the sigma
models with almost complex manifolds as a target space as well as
those associated with $(p,0)$ fermionic multiplets for which the
supersymmetry algebra closes on-shell \cite{phgpb,hw}. Other models
of interest that we have not described here are those with $(p, 2)$,
$p=2,4$, and (4,4) supersymmetry. All the above models can be
described using (1,0) superfields. This method has been used before,
see \cite{gppt,gppt2,gppt3}. This means that the action of such models
can be written in terms of (1,0) superfields and the additional
supersymmetries can be implemented by requiring invariance of the
action under additional suitable transformations. The (2,2) and (4,4)
supersymmetric gauge theories have been described using other methods
in \cite{witten} and \cite{crisaru}.

The gauge theories coupled to sigma models which we have described
with $(p,1)$ supersymmetry have soliton type of bounds in addition to
the vortex type of bounds that we have described.  For the former
bounds the energy of these models can be written as a sum of squares
and a topological term. This is very similar to bounds of (ungauged)
sigma models \cite{apt} and so we have not described them here. It
would be of interest to investigate the solutions of the vortex
equations we have presented for different types of moment maps. It may
be that for a suitable choice, the vortex equations can be solved
exactly.

 \vskip 0.5cm

\noindent{\bf Acknowledgements}
  We would like to thank  D. Tong and  P. Townsend
for useful discussions. One of us GP thanks the theory group
at CERN for hospitality. This work is partly
 supported by the PPARC grant
   PPA/G/S/1998/00613. GP is
supported by a University
Research Fellowship
from the Royal Society.

\providecommand{\href}[2]{#2}\begingroup\raggedright\endgroup

\end{document}